\documentclass{article}
\pdfoutput=1

\usepackage[utf8]{inputenc}
\usepackage[english]{babel}

\usepackage{authblk}

\usepackage{amsmath}
\usepackage{graphicx}
\usepackage{parskip}
\usepackage{siunitx}

\usepackage{algorithm}
\usepackage{algpseudocode}
\usepackage{filecontents}
\usepackage{listings}
\usepackage{xr}

\usepackage{caption}
\usepackage{etoolbox}

\usepackage{csquotes}
\MakeOuterQuote{"}

\makeatletter
\pretocmd\lst@makecaption{\noindent{\rule{\linewidth}{1pt}}}{}{}
\makeatother

\DeclareMathOperator\erf{erf}

\usepackage[style=apa]{biblatex}
\title{Hidden assumptions of integer ratio analyses in bioacoustics and music}

\author[1]{Yannick Jadoul}
\author[2]{Tommaso Tufarelli}
\author[1]{Chloé Coissac}
\author[3]{Marco Gamba}
\author[1,4,5]{Andrea Ravignani}
\affil[1]{Department of Human Neurosciences, Sapienza University of Rome, Rome, Italy}
\affil[2]{Independent researcher}
\affil[3]{Department of Life Sciences and Systems Biology, University of Turin, Turin, Italy}
\affil[4]{Center for Music in the Brain, Department of Clinical Medicine, Aarhus University, Aarhus,
Denmark}
\affil[5]{Research Center of Neuroscience ``CRiN-Daniel Bovet'', Sapienza University of Rome, Rome, Italy}
\affil[ ]{Corresponding author: Yannick.Jadoul@uniroma1.it}

\date{}

\begin{document}

\maketitle

\begin{abstract}
\noindent 
Rhythm is ubiquitous in human culture and in nature, but hard to capture in all its complexity. A key dimension of rhythm, integer ratio categories occur when the relationship between temporal intervals can be expressed as small-integer ratios. Recent work has found integer ratio categories in most human musical cultures and some animal species’ vocalizations or behavioral displays. But biological systems are noisy, and empirically measured intervals rarely form an exact small-integer ratio. Here, we mathematically assess whether the leading integer ratio analysis method makes valid statistical and biological assumptions. In particular, we (1) make the temporal properties of empirical ratios explicit, both in general and for the typical use in the literature; (2) show how the choice of ratio formula affects the probability distribution of rhythm ratios and ensuing statistical results; (3) guide the reader to carefully consider the assumptions and null hypotheses of the statistical analysis; (4) present a comprehensive methodology to statistically test integer ratios for any null hypothesis of choice. Our observations have implications for both past and future research in music cognition and animal behavior: They suggest how to interpret past findings and provide tools to choose the correct null hypotheses in future empirical work.
\end{abstract}

\textbf{Keywords:} categorical rhythm, animal behavior, vocalization, timing, beat, meter, statistical assumptions

\newpage

\section{Introduction: Integer ratios as a measure of rhythmic structure}
Quantifying temporal patterns is a crucial step toward a better understanding of rhythm and its origins. One common pattern in human music are intervals whose durations conform to small-integer ratios; for example, the durations of a quarter note and an eight note are related in a 2:1 ratio \parencite{roeske_categorical_2020}. Small-integer ratios are ubiquitous across individuals and cultures, and appear to be a striking universal feature of human music and rhythm \parencite{ravignani_musical_2016, jacoby_integer_2017, jacoby_commonality_2024}. Humans both perceive and produce small-integer ratios in temporal sequences; a few hypotheses exist about the underlying mechanisms \parencite{ravignani_why_2018}. Investigating the presence of rhythmic categories conforming to small-integer ratios is one way of measuring temporal structure in behavior, and perhaps even inferring internal mechanisms of the biological agents producing those behaviors \parencite{kotz_evolution_2018, de_gregorio_categorical_2021, hersh_robust_2023}.

In the search for a better understanding of the evolutionary origins of this fundamental building block of rhythm perception and production, recent studies have started quantifying the degree to which small-integer ratios occur across animals. Small-integer ratios are not only found in human music \parencite{jacoby_integer_2017, roeske_categorical_2020, jacoby_commonality_2024}, but also in a broad range of non-human animal behaviors: Many species show high rates of isochrony (1:1 ratio) in their vocal displays (e.g., several gibbon species, \cite{raimondi_isochrony_2023, de_gregorio_isochronous_2024, ma_small_2024}; orangutans, \cite{lameira_recursive_2024}; rock hyraxes, \cite{demartsev_male_2023}; zebra finches, \cite{roeske_categorical_2020}). Several other species show different rhythmic categories (e.g., thrush nightingales, \cite{roeske_categorical_2020}; indri, \cite{de_gregorio_categorical_2021}; Australian pied butcherbirds, \cite{xing_syntactic_2022}). In other studies, while no clear rhythmic categories are found in vocalizations \parencite{anichini_measuring_2023}, rhythmic categories are present in non-vocal displays \parencite[e.g., harbor seals' flipper slapping,][]{kocsis_harbour_2024} or behaviour \parencite[e.g., horse gait,][]{laffi_rhythm_2025,laffi_rhythmic_2025}.

Here, we dissect a recent approach for quantifying and testing the presence of rhythmic categories at small-integer ratios in empirical data, pioneered in non-human animals by \textcite{roeske_categorical_2020}. In the following section (Section~\ref{sec:background}), we first shortly outline this approach. Next, we present a summary of our findings using a minimum of mathematical jargon or formulas (Section~\ref{sec:results-summary}). The unconcerned reader should read Section~\ref{sec:empirical-example} right after Section~\ref{sec:results-summary}, and skip Sections~\ref{sec:mathematical-properties}, \ref{sec:statistical-inference}, and \ref{sec:alternative-null-hypothesis}, where we go into full detail about the mathematical and statistical results summarized before. To guide the interpretation of the presented equations, Table~\ref{tab:notation} provides a summary of the notation used. Based on previous empirical data, we demonstrate the presented formulas with a fully worked-out example in Section~\ref{sec:empirical-example}. Finally, in Section~\ref{sec:conclusion}, we present our results and takeaway message for the empirical scientist intending to apply the approach, refraining from technical details.

\section{Background: How integer ratios are typically measured}
\label{sec:background}
\begin{figure}[!h]
	\centering
	\includegraphics[width=\linewidth]{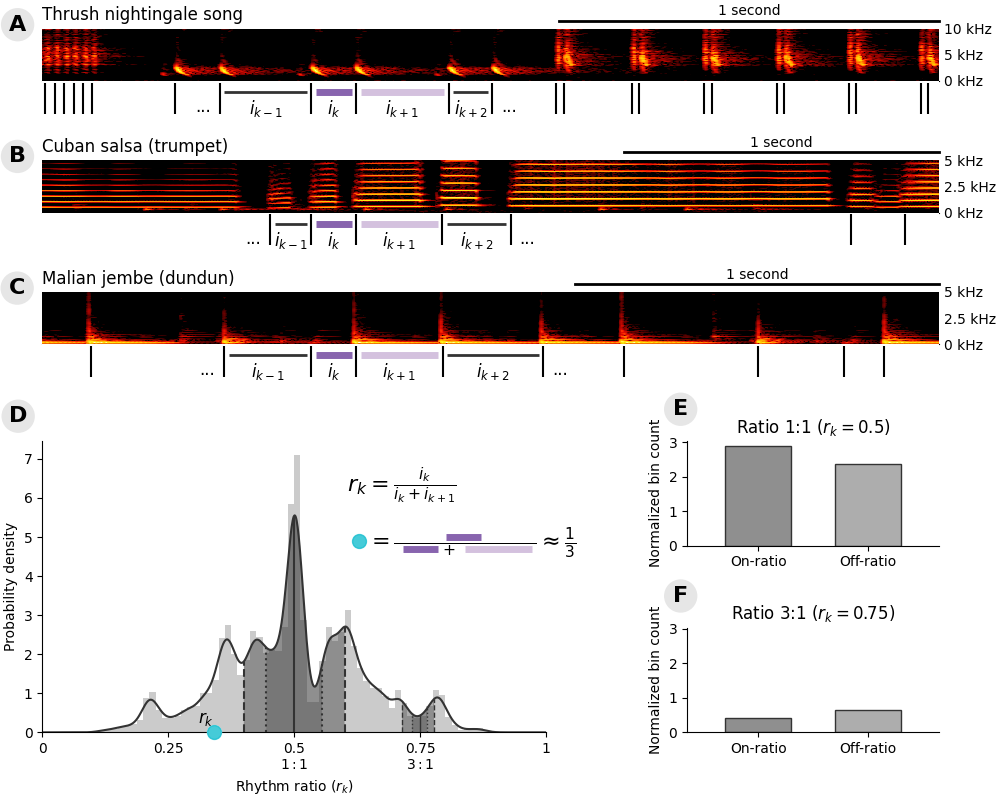}
	\caption{The rhythmic analysis of an audio recording starts by extracting the events and intervals. For example, these events could be the onset of syllables in thrush nightingale's song (\textbf{A}; \cite{roeske_birdsong_2020}), the onset of the musical notes played by a a trumpet (\textbf{B}; \cite{roeske_musical_2020, clayton_interpersonal_2018}), or the drumbeats of a dundun (\textbf{C}; \cite{roeske_musical_2020, clayton_interpersonal_2018}). The rhythm ratio formula (\textbf{D}, equation~\ref{eq:r_k}) is applied to each pair of adjacent intervals in a temporal sequence (\textbf{A}, \textbf{B}, \textbf{C}), resulting in a distribution of rhythm ratios (\textbf{D}). The region around a specific integer ratio is split into several parts: an on-ratio and two off-ratio bins (dark and light shading). The fraction of rhythm ratios $r_k$ contained in each bin is then normalized by the total number of ratios and by the total bin width. The resulting values are empirical estimates of the probability density per bin (for example, 1:1 and 3:1, \textbf{D} and \textbf{E}), which are then compared using a statistical test. Note that we have combined both off-integer bins into a single normalized bin count, following \textcite{de_gregorio_categorical_2021} and others. Other studies keep both normalized bin counts separated \parencite{roeske_categorical_2020}. In the context of the current study, the difference between these approaches is irrelevant.}
	\label{fig:example-bins}
\end{figure}

How to investigate the presence of integer ratios in \emph{empirical} temporal sequences is not a trivial problem. Any measurement on a biological system produces noisy data. In the temporal domain, such noise will make it challenging to detect small-integer ratios \textit{precisely}. Hence, we need a methodological framework that allows us to \emph{statistically} assess whether the observed ratios are concentrated around small-integer ratios. The approach presented here starts from an observation of a potentially rhythmic temporal sequence (e.g., the notes in a piece of music or the onset of calls in an animal display; see Figure~\ref{fig:example-bins}). To test for rhythmicity in such a sequence, we measure the intervals between events, and investigate whether the ratios between subsequent intervals approximate a small-integer ratio more often than expected. This then allows us to determine whether or not those rhythmic patterns are prominent aspects of that temporal sequence.

Below, we examine and generalize the approach by \textcite{roeske_categorical_2020}, which analyses the ratio between adjacent intervals in temporal sequences. Limiting the analysis to only adjacent intervals has the practical advantage of ignoring a slow drift in tempo throughout a sequence. Moreover, while such analysis may fail to explicitly capture other rhythmic patterns, the presence of small-integer ratios implies non-random relationships between adjacent intervals. Measuring ratios between adjacent intervals also captures the local rhythmic structure found in human perception and production \parencite{jacoby_integer_2017}.

The \emph{rhythm ratio} $r_k$ quantifies the relationship between two adjacent intervals considering the ratio of one interval to their joint duration \parencite{roeske_categorical_2020}. More precisely, given a sequence of temporal intervals\footnote{In some studies \parencite[e.g.,][]{de_gregorio_categorical_2021}, intervals are denoted as $t_k$. To avoid potential confusion with points in time, we here stick to the $i_k$ notation.} $i_1, i_2, \ldots, i_k, \ldots, i_n > 0$, $r_k$ is calculated as follows:

\begin{align}
r_k = \frac{i_k}{i_k + i_{k+1}} \label{eq:r_k}
\end{align}

The resulting ratio $r_k$ has several straightforward and enticing properties:
\begin{enumerate}
	\item \emph{Only the relative duration of intervals matters}: \\
	      E.g., two adjacent intervals of \qty{1}{\s} and \qty{2}{\s} result in a rhythm ratio of $r_k = \frac{\SI{1}{\s}}{\SI{1}{\s} + \SI{2}{\s}} = \frac{1}{3} = 0.333\ldots$; two much shorter intervals of \qty{25}{\ms} and \qty{50}{\ms} with the same relative durations have the exact same ratio: $\frac{\qty{25}{\ms}}{\qty{25}{\ms} + \qty{50}{\ms}} = \frac{1}{3} = 0.333\ldots$
	\item \emph{Ratios $r_k$ are bounded between 0 and 1}: \\
	      E.g., for two intervals of extremely different durations, \qty{1}{\ms} and \qty{1}{\s} (i.e., \qty{1000}{\ms}), the $r_k$ ratio is almost 0 ($\frac{\qty{1}{\ms}}{\qty{1}{\ms} + \qty{1000}{\ms}} = 0.000999\ldots$) or almost 1 ($\frac{\qty{1000}{\ms}}{\qty{1}{\ms} + \qty{1000}{\ms}} = 0.999000\ldots$), depending on the intervals' order.
	\item \emph{Inverse integer ratios, $m:n$ and $n:m$, sum to 1 and are symmetric around $1:1$}: \\
		  E.g., the rhythm ratio corresponding to integer ratio $1:2$ is $\frac{1}{1 + 2} = 0.333\ldots$, and the rhythm ratio corresponding to integer ratio $2:1$ is $r_k = \frac{2}{2 + 1} = 0.666\ldots = 1 - \frac{1}{1 + 2}$; so, integer ratios $1:2$ and $2:1$ have $r_k$ values that add up to 1 and are symmetric around 0.5 (corresponding to $1:1$, or isochrony).
\end{enumerate}

Note that for a simpler formula, such as the direct fraction of two intervals $q_k = \frac{i_{k+1}}{i_k}$ \parencite[e.g.,][]{ravignani_musical_2016}, properties 2 and 3 do not hold: Following this formula to quantify ratios, the $q$ values for integer ratios $2:1$, $3:1$, and $4:1$ lie very close together between 0.25 and 0.5, while the inverse integer ratios $1:2$, $1:3$, and $1:4$ are spread out between 2 and 4. Conversely, the corresponding $r_k$ values lie equally close to each other, respectively ranging $0.666\ldots$ to $0.8$ and between $0.2$ to $0.333\ldots$. However, there could be other, unexplored ratio measures which might feature these same or other convenient properties (see below, Sections~\ref{sec:results-summary} and \ref{sec:alternative-null-hypothesis}).

After calculating the rhythm ratios in empirical interval sequences (Figures~\ref{fig:example-bins}A \& \ref{fig:example-bins}B), the resulting distribution of rhythm ratios can be inspected for clusters around \emph{small-integer} ratios (i.e., peaks in the density distribution; see Figure~\ref{fig:example-bins}C). The approach taken by recent research is to count the number of rhythm ratios falling into several bins around small-integer ratios (see Figures~\ref{fig:example-bins}C, \ref{fig:example-bins}D, \ref{fig:example-bins}E). In short, to statistically test the presence of an integer ratio peak in the data, all points within a region of interest around that ratio are assigned to one of two groups: on-integer ratio and off-integer ratio values; the former data points fall in the vicinity of the integer ratio, while the latter away from it \parencite[for details, see][]{roeske_categorical_2020, de_gregorio_categorical_2021}.

These counts in different bins around an integer ratio can then be compared using a statistical method, through e.g., bootstrapping \parencite{roeske_categorical_2020}, a Wilcoxon signed-rank test \parencite{de_gregorio_categorical_2021, ma_small_2024}, or generalized linear mixed models \parencite{raimondi_isochrony_2023, lameira_recursive_2024}. Crucially, in most cases, the on- and off-integer bins proposed by \textcite{roeske_categorical_2020} do not have the same width; hence, their respective counts are normalized by dividing by each bin's width before statistically comparing them. If a bin of width $w$ contains $m$ rhythm ratio values out of a total of $N$ values, this normalized count equals $\frac{m}{N \cdot w}$ \parencite{roeske_categorical_2020}. In other words, one divides the relative proportion of observations in that bin by the width of the bin. This normalization intends to allow for a fair comparison between proportions of observations that fall into broad bins, which by pure chance are likely to contain many observations, with those that fall into narrow bins.

Roughly speaking, regardless of the test one decides to employ, the test's null hypothesis will be that there is no difference between the number of on-peak ratios and the number of off-peak ratios. A significant result associated with a higher number of on-peak $r_k$ values for a certain small-integer ratio will, therefore, indicate a discrepancy sufficient to suggest that that particular rhythmic category is present. As we demonstrate below, this view is actually simplistic: Slight variations in the calculation of $r_k$ and the bins' normalization may correspond to non-trivial deviations in the resulting statistical analysis.

This approach has proven convenient and practical to investigate the presence of integer ratios in temporal sequence data. Three main methodological questions with practical implications that seem to have remained unexplored:
\begin{itemize}
	\item Many formulas could capture the ratio between two intervals. What are the mathematical properties of the rhythm ratio $r_k$ formula? (Section~\ref{sec:mathematical-properties})
	\item Statistical tests make mathematical assumptions about the data and its distribution. How does the choice of formula affect the probability distribution of rhythm ratios and the ensuing statistical results? (Section~\ref{sec:statistical-inference})
	\item Assumptions which do not match reality or inappropriate null hypotheses might render a statistical result useless. Are there alternative ways of transforming intervals to visualize and test integer ratios, making different assumptions or offering different advantages? (Section~\ref{sec:alternative-null-hypothesis})
\end{itemize}

\section{Summary of results: The importance of the right formula and null hypothesis}
\label{sec:results-summary}

\emph{Many formulas could capture the ratio between two intervals. What are the mathematical properties of the rhythm ratio $r_k$ formula?}

The $r_k$ formula presented above (equation~\ref{eq:r_k}) is only one way of measuring the relationship between two intervals, as there are many other formulas that could quantify the relationship between $i_k$ and $i_{k+1}$ in a single value. One property of the $r_k$ formula is nevertheless particularly noteworthy: The resulting $r_k$ value is not affected by the overall tempo of a temporal sequence. This is usually a desirable property when investigating rhythmic patterns and the presence of small integer ratios. Further analysis shows that there are indeed infinite alternative ratio formulas with this same property; for any such formula to be tempo-independent, it can only depend on the fraction of the two intervals.

However, the choice of a ratio formula also influences the resulting distribution of the ratios. For example -- as noted in Section~\ref{sec:background} -- using the straightforward fraction of two intervals ($\frac{i_{k+1}}{i_k}$) introduces asymmetry. Consequently, the chosen rhythm ratio formula plays a crucial role in the visualization and analysis of a temporal sequence.

\begin{figure}[!h]
	\centering
	\includegraphics[width=\linewidth]{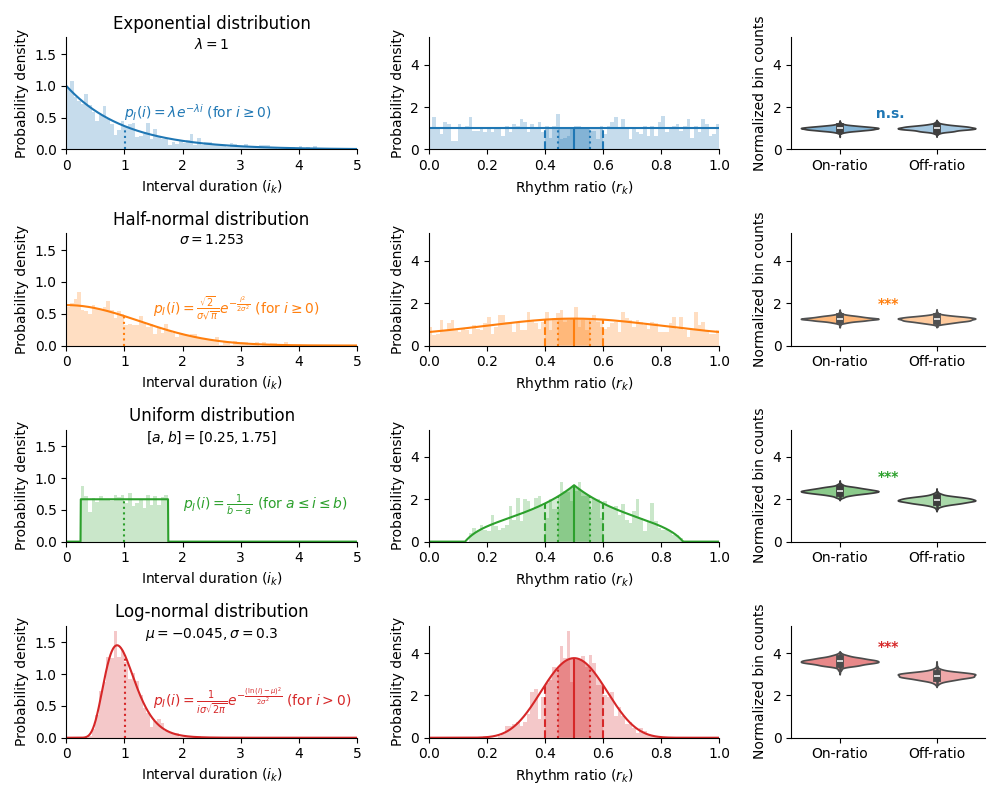}
	\caption{Four example probability distributions of interval durations $i_k$ (left column) all result in differently-shaped probability distributions of rhythm ratios $r_k$ (middle column). The theoretically derived probability density functions (full lines) match the distributions of $r_k$ in a randomly sampled sequence of 1000 intervals (histograms). A half-normal, a uniform, and a log-normal distribution of intervals all result in non-uniform distributions of $r_k$ with a higher density around $r_k = 0.5$ (ratio 1:1). Normalizing bin counts by bin width results significantly more often in a higher normalized count of on-1:1 ratio $r_k$ (right column, 1000 sequences of 1000 intervals, Wilcoxon signed-rank test; half-normal distribution: $T = 209379$, $p < 10^{-5}$; uniform distribution: $T = 515$, $p < 10^{-163}$; log-normal distribution: $T = 98$, $p < 10^{-164}$). For a exponential distribution of intervals (i.e., a Poisson process), the normalized on- and off-ratio counts for 1:1 do not differ significantly (Wilcoxon signed-rank test; $T = 235643$, $p \approx 0.11$).}
	\label{fig:ratio-distributions}
\end{figure}

We can investigate the mathematical properties of the $r_k$ ratio by exploring which $r_k$ values are generated by different random processes. Our derivations show how the random distribution of a sequence's interval durations ($i_k$) determines the resulting distribution of $r_k$ values (see Figure~\ref{fig:ratio-distributions} and equation~\ref{eq:p_r}). We present one particular mathematical process for generating temporal events and intervals that provides a lot of insight into the $r_k$ values: the Poisson point process. The expected distribution of $r_k$ ratios resulting from such a highly random process is a uniform distribution; i.e., for a Poisson process, independent of its rate parameter, the values of $r_k$ will equally likely fall anywhere between 0 and 1. As probability distributions and statistics are strongly connected, this means that the choice of ratio formula also has an influence on any subsequent statistical inference.

\emph{Statistical tests make mathematical assumptions about the data and its distribution. How does the choice of formula affect the probability distribution of rhythm ratios and the ensuing statistical results?}

A key feature of statistical analyses is the assumption of a null hypothesis, formally specifying the statistical properties of the data if the studied effect would not exist. Most studies performing statistical tests on the rhythm ratios $r_k$ count the number of ratios that fall close to a small-integer ratio. As described earlier (Section~\ref{sec:background}), these counts are then normalized by bin width and statistically compared. Crucially, the count of a bin that is wider will be divided by a higher normalization factor: This means that under the null hypothesis, the number of $r_k$ values in each bin is proportional to a bin's width. This null hypothesis, resulting from the normalization by a bin's width, is equivalent to assuming that $r_k$ values are randomly and uniformly distributed between 0 and 1.

This brings us to a key result: \emph{Previous studies have compared their empirical $r_k$ distributions against an implicit null hypothesis.} This implicit null hypothesis being tested is that the data have been generated by a Poisson process. Any significant results based on normalized bin counts \parencite[following][]{roeske_categorical_2020, de_gregorio_categorical_2021} statistically show that the data was not generated by a random Poisson process. Even if this null hypothesis was not made explicit, statistical significance shows that the animal vocalizations feature rhythmic categories different from one particular type of random timing. One important caveat follows, as a Poisson process generates highly random temporal sequences: This implicitly assumed null hypothesis represents a relatively low bar for testing the presence of rhythm categories, and is not necessarily an indication of extraordinary rhythmic capabilities. As Figure~\ref{fig:ratio-distributions} shows, any limit on the possible range of interval durations produced by an animal (e.g., because of biomechanical constraints) will also cause $r_k$ values to be significantly different from a uniform distribution.

\emph{Assumptions which do not match reality or inappropriate null hypotheses might render a statistical result useless. Are there alternative ways of transforming intervals to visualize and test integer ratios, making different assumptions or offering different advantages?}

During statistical analysis, the distribution of observed rhythm ratios is usually compared to a random baseline distribution, representing the null hypothesis. As we show, this baseline is often a uniform distribution of ratios, which corresponds to a Poisson process and an exponential distribution of interval durations (Figure~\ref{fig:ratio-distributions}). However, in certain cases, one might want to test the empirical data against a different null hypothesis. For example, if previous observations or theoretical considerations make it likely that an animal's produced interval durations fall between a minimum and a maximum value, a uniform distribution over this range can be a natural null hypothesis. This case corresponds to the simulated null distributions and Kolmogorov--Smirnov tests in a handful of previous studies \parencite{de_gregorio_categorical_2021, xing_syntactic_2022, anichini_measuring_2023, demartsev_male_2023}. Similarly, when a behavior is hypothesized to produce log-normally distributed intervals, one might want to test whether certain integer ratios occur more frequently wrt. a log-normal distribution of intervals.

To test empirical data against a different null hypothesis, there are two options: The first option is to use a different rhythm ratio formula (instead of $r_k$; equation~\ref{eq:r_k}), which would result in a uniform distribution of \emph{rhythm ratios} for the baseline distribution of intervals (see equations~\ref{eq:f_+} and \ref{eq:f_-} below). The second, often more practical option is to weight the data according to its expected probability. For example, following the mathematical derivations we present below, we can calculate the probability that a rhythm ratio $r_k$ falls between 0.4 and 0.5 or between 0.3 and 0.4 if the interval durations are uniformly distributed. To statistically test whether rhythm ratios fall more than expected by chance in one of these two bins, the counts should be normalized by the expected probability of each bin (i.e., the area under the null distribution's curve), instead of the bin widths. A comparison between these normalized bin counts will correctly take into account the underlying null hypothesis of the intervals' distribution. Section~\ref{sec:empirical-example} demonstrates the latter approach by applying it to the same empirical data analyzed by \textcite{roeske_categorical_2020}.

\section{Mathematical properties of rhythm ratios}
\label{sec:mathematical-properties}

\begin{table}[h]
	\centering
	\begin{tabular}{ll}
		Notation & \\
		\hline
		$m:n$ & integer ratio between two intervals \\
		$i_k$ & empirical interval in a sequence \\
		$q_k$ & empirical fraction between two interval durations in a sequence \\
		$r_k$ & empirical rhythm ratio in a sequence \parencite{roeske_categorical_2020} \\
		$r_{m:n}$ & rhythm ratio of a perfect integer ratio \\
		\hline
		$i_1$, $i_2$ & any two intervals \\
		$q$ & fraction between two interval durations \\
		$r$ & standard rhythm ratio of two intervals \parencite{roeske_categorical_2020} \\
		$s$ & any alternative rhythm ratio of two intervals \\
		$F$ & any transformation function from $i_1, i_2$ to $s$ \\
		$f$ & any transformation function from $q$ to $s$ \\
		$p_I$ & probability density functions of intervals \\
		$p_Q$, $p_R$, $p_S$ & probability density functions of $q$, $r$, $s$ \\
		$P_Q$, $P_R$, $P_S$ & cumulative probability functions of $q$, $r$, $s$ \\
		\hline
		$w$ & width of a $r_k$ ratios bin \\
		$\hat{w}_{I, u, v}$ & normalization factor for a bin $[u, v]$ for intervals distributed as $p_I$ \\
		\hline
		$c$ & auxiliary variable representing a scaling factor \\
		$t$ & auxiliary variable being integrated \\
		\hline
		$\lambda$ & Rate of the Poisson point process \\
		$a$, $b$ & The minimum and maximum of a uniform interval distribution $p_I$ \\
		$\mu$, $\sigma$ & Logarithm of location and scale of the log-normal distribution \\
		\hline
	\end{tabular}
	\caption{Overview of the mathematical notation used in derivations and formulas below.}
	\label{tab:notation}
\end{table}

\subsection{Scale-invariant rhythm ratio formulas}

The three properties of $r_k$ mentioned in Section~\ref{sec:background} are straightforward to show mathematically:

\begin{enumerate}
	\item \emph{Only the relative duration of intervals matters}: \\
	Scaling $i_k, i_{k+1}$ by a factor $c > 0$, the resulting rhythm ratio remains the same, as $r'_k = \frac{c \cdot i_k}{c \cdot i_k + c \cdot i_{k+1}} = \frac{i_k}{i_k + i_{k+1}} = r_k$.
	\item \emph{Ratios $r_k$ are bounded between 0 and 1}: \\
	For any interval durations $i_k, i_{k+1} > 0$, the value of $r_k$ is bounded by $0 = \frac{0}{0 + i_{k+1}} < r_k < \frac{i_k}{i_k + 0} = 1$.
	\item \emph{Inverse integer ratios, $m:n$ and $n:m$, sum to 1 and are symmetric around $1:1$}: \\
	Two equally sized intervals will have a ratio $r_{1:1} = \frac{1}{1 + 1} = \frac{1}{2} = 0.5$ \\
	Two inverse ratio rhythms, $r_{m:n}$ and $r_{n:m}$, will add up to 1, as $r_{n:m} + r_{m:n} = \frac{n}{n + m} + \frac{m}{m + n} = 1$. It follows that those two ratios will be symmetrically distributed around $0.5$, as $\frac{1}{2} \left(r_{n:m} + r_{m:n}\right) = \frac{1}{2} = r_{1:1}$.
\end{enumerate}

In the context of integer ratios, it appears natural that the underlying tempo should not affect how the relative relationship between intervals is quantified; in other words, we require any rhythm ratio formula to be scale-invariant.

Consider any rhythm transformation function $F(i_1, i_2),$ with interval durations $i_1, i_2>0$. If we impose the scale invariance condition, then for any $c > 0$,
\begin{align}
    F(c\, i_1, c\, i_2) = F(i_1, i_2)
\end{align}

We can show that $F$ must only depend on the ratio $q = \frac{i_2}{i_1}$, i.e. $F(i_2, i_1) = f(q)$ for some function $f$. As proof, choose $c = \frac{1}{i_1}$, such that $c\, i_1 = 1$ and $c\, i_2 = \frac{i_2}{i_1} \equiv q$. Then $F(i_1, i_2) = F(c\, i_1, c\, i_2) = F(1, q),$
such that it is easy to identify 
\begin{align}
    f(q) = F(1, q)
\end{align}

In conclusion, we can reduce any scale-invariant quantification of relative rhythm (between two intervals) to a single-variable function $f(q)$ that is only dependent on the fraction $q = \frac{i_2}{i_1}$. Conversely, any function $f(q)$ of $q = \frac{i_2}{i_1}$ is scale-invariant by construction, since the ratio does not change when both intervals are scaled by the same constant: $\frac{c\, i_2}{c\, i_1} = \frac{i_2}{i_1} = q$.

The rhythm ratio $r_k$ (equation \ref{eq:r_k}; \cite{roeske_categorical_2020}) fulfills this property, the explicit form of the corresponding $f$ being $f(q) = \frac{1}{1 + q} = \frac{i_1}{i_1 + i_2}$. This neatly fits with the observation that a sequence of $n$ intervals $i_k$ has only $n - 1$ $r_k$ values. This single degree of freedom that gets lost when calculating a temporal sequence's $r_k$ values, directly corresponds to the average tempo -- or equivalently, the mean $i_k$ value.

\subsection{The probability distribution of rhythm ratios}
In a perfectly isochronous sequence, all intervals are the same size (i.e., ratio 1:1). In this case, all the sequence's $q_k$ and $r_k$ values will be equal, since $q_k = \frac{1}{1} = 1$ and $r_k = \frac{1}{1 + 1} = 0.5$. Similarly, for other temporal sequences built using only a few exact interval durations, the resulting distribution of $q_k$ and $r_k$ will contain a limited number of unique values. However, it is more interesting to consider the case when intervals can vary continuously and are randomly sampled from a certain probability distribution.

A first step towards understanding the behavior of rhythm ratios is calculating the probability distribution of the ratio $q$. Assuming two intervals $i_1, i_2 > 0$ distributed according to a joint probability density $p_I(i_1, i_2),$ we can calculate the resulting probability density function of the ratio $q$, $p_Q(q)$. To find the two-dimensional probability distribution over $i_1$ and $q = \frac{i_2}{i_1}$, we replace variable $i_2$ and write it in function of $q$, setting $i_2 = q\, i_1$: $p_{I'}(i_1, q) = i_1\, p_I(i_1, q\, i_1)$. Subsequently, we calculate the marginal probability distribution over \emph{only} $q$, rename $i_1$ as $t$, and integrate over all possible values of $t$. Performing this change of variables from $(i_1, i_2) \to (t, q\, t),$ and integrating over $t$, simplifies to the following equation \parencite{springer_algebra_1979, ross_first_2019}:

\begin{align}
    p_Q(q) &= \int_0^\infty t\, p_I(t, q\, t)\, dt \label{eq:p_q_joint}
\end{align}

Additionally, if $i_1$ and $i_2$ are statistically independent, $p_I(i_1, i_2) = p_I(i_1)\, p_I(i_2)$, and equation \ref{eq:p_q_joint} reduces to
\begin{align}
    p_Q(q) = \int_0^\infty t\, p_I(t)\, p_I(q\, t)\, dt \label{eq:p_q}
\end{align}

Given $p_Q(q)$, we can also determine the probability density $p_{S}$ of any rhythm ratio $s$ given its transformation $f(q) = s$, if $f$ has an inverse $f^{-1}$. If so, another change of variables in the probability density function, $q \to s = f(q)$, shows that
\begin{align}
    p_{S}(s) = p_Q\left(f^{-1}(s)\right) \left|\frac{d}{ds}f^{-1}(s)\right| \label{eq:p_s}
\end{align}

Altogether, for the rhythm ratio $r = \frac{i_1}{i_1 + i_2}$, where $f(q) = \frac{1}{1 + q}$, $f^{-1}(r) = \frac{1 - r}{r}$,
\begin{align}
    p_R(r) &= \frac{1}{r^2} \int_0^\infty t\, p_I(t)\, p_I\left(t\, \frac{1 - r}{r}\right) dt \label{eq:p_r}
\end{align}

So, given a probability distribution of interval durations $i_k$, with density $p_I$, equation~\ref{eq:p_r} describes the resulting probability density of resulting ratios $r_k$.

\subsection{Rhythm ratios of a Poisson point process}
One relevant stochastic process to consider for generating a temporal sequence is a Poisson point process. A uniform Poisson process with rate $\lambda$ generates events in time which are completely independent from each other and are separated by exponentially distributed intervals; i.e.,
\begin{align}
    p_I(i) = \lambda e^{-\lambda\, i} \label{eq:p_i_poisson}
\end{align}

where $1/\lambda$ is the average waiting time for an event to occur.

A Poisson point process is often considered a `highly random' way of generating a temporal sequence. The events in a Poisson process happen  independently from each other, and the process is "memoryless" in the sense that the likelihood of an event only depends on the system's current state and not its past history. Poisson processes are routinely used to mathematically model, e.g., the radioactive decay of individual atoms.

Following equations \ref{eq:p_q} and \ref{eq:p_r} above, it can be shown that the probability distribution of the ratios $q$ and $r$ resulting from two independent and exponentially distributed intervals equals
\begin{align}
    p_Q(q) &= \frac{1}{(1 + q)^2} \label{eq:p_q_poisson}
\end{align}

\begin{align}
    p_R(r) &= 1
\end{align}

In other words, independently of a uniform rate $\lambda$, the rhythm ratios $r_k = \frac{i_k}{i_k + i_{k+1}}$ \parencite[as proposed by][]{roeske_categorical_2020} of intervals generated by a Poisson process follow a uniform distribution on $[0, 1]$. The Poisson process's rate $\lambda$ does not appear in the equations $p_Q(q)$ and $p_R(r)$ because its value corresponds to the average tempo. As shown earlier, $q$, $r$, and any other scale-invariant rhythm ratio capture the rhythmic relationship between two interval durations independently of the underlying tempo.

This is an attractive property of the $r_k$ rhythm ratio, since the Poisson point process produces highly random temporal sequences and is thus a good baseline to which other sequences can be compared. One could even argue that in many ways it is a maximally random process for temporal sequences, as the exponential distribution is the maximum entropy distribution on $[0, \infty)$ with a fixed mean of $\frac{1}{\lambda}$. In plainer terms, this means that a Poisson process generates the least predictable sequence of intervals (with unbounded length) and a fixed mean. Correspondingly, the uniform distribution of $r_k$ ratios is the distribution which achieves maximum entropy on a bounded interval (such as $[0, 1]$). This connection between a Poisson process and a uniform distribution of $r_k$ values, both maximum entropy distributions, provides a strong theoretical justification for the use of the $r_k$ rhythm ratio formula.

\section{Statistical inference of rhythm ratios}
\label{sec:statistical-inference}
As shown above, the widely-used rhythm ratio $r_k = \frac{i_k}{i_k + i_{k+1}}$ results in a uniform distribution over $[0, 1]$ for a Poisson point process (i.e., when $i_k$ and $i_{k+1}$ follow an exponential distribution with the same rate $\lambda$; see exponential distribution of intervals in Figure~\ref{fig:ratio-distributions}). This mathematical link has consequences for statistical analyses based on this ratio $r_k$.

Many statistical approaches compare an empirical sample of ratios $r_k$ to a uniform distribution, implicitly or explicitly. Any such statistical analysis implicitly assumes the null hypothesis that the data was generated by a homogeneous Poisson point process. This holds in particular for the typical approach where calculated rhythm ratios are binned and those counts are normalized by the bin widths \parencite[see Section~\ref{sec:background} and Figure~\ref{fig:example-bins}; e.g.,][]{roeske_categorical_2020, de_gregorio_categorical_2021}. Normalizing bin counts by their width implies a baseline distribution where the expected number of points in each bin is directly proportional to its width. In other words, this division normalizes bin counts with respect to a uniform distribution over $[0, 1]$. So a statistical comparison between normalized bin counts de facto tests the implicit null hypothesis that the (binned) empirical rhythm ratios have been generated by a Poisson process. This goes for most studies which have relied on the $r_k$ ratio formula: Almost all of past research has tested, without explicit mention of it, whether the empirical data is clustered around certain small-integer ratios \emph{significantly more than a Poisson point process}.

Many studies has assumed a Poisson point process as the null hypothesis. This is an important nuance, since a Poisson process produces highly random rhythmic sequences \textit{with no limitation on the interval length}. Therefore, a Poisson process represents a relatively weak null hypothesis to be rejected. Depending on the rhythmic sequences or animal displays being investigated, studies might want to test a stricter null hypothesis, perhaps tailored to the species or task at hand. For example, intervals $i_k$ sampled independently from a uniform distribution, with a maximum interval length, will result in ratios $r_k$ which do \emph{not} follow a uniform distribution \parencite[see simulated null distributions and Kolmogorov--Smirnov tests by][]{de_gregorio_categorical_2021, xing_syntactic_2022, anichini_measuring_2023, demartsev_male_2023}. In fact, further analyses and computational simulations show that this distribution has a peak around integer ratio 1:1, causing the null hypothesis to be rejected more often than by chance (see Figure~\ref{fig:ratio-distributions}). This is a correct rejection; i.e., the intervals were \emph{not} generated by a Poisson process. However, this null hypothesis might not be the one authors should be trying to reject, since the resulting peak at 1:1 is still due to a random process (albeit a different one than the Poisson process).

\section{Choosing a different null hypothesis}
\label{sec:alternative-null-hypothesis}

\subsection{Alternative null hypotheses}
As illustrated above, the use of the rhythm ratio $r_k = \frac{i_k}{i_k + i_{k+1}}$ often goes hand in hand with one key assumption: in statistical tests, empirical data is compared to a Poisson process as null hypothesis. However, depending on the research question and its context, a Poisson process might not be the most appropriate choice. For example, a biological system may feature physiological constraints such as a minimum duration between two movements or a maximum lung capacity. These constraints limit the range of the interval distribution by imposing minimum and maximum durations, while the Poisson has a non-zero probability of generating extremely small or extremely large intervals. Temporal sequences generated according to such additional constraints are, mathematically speaking, less random than a Poisson process, and will exhibit different statistical properties.

Using the above mathematical insights, we can however test different types of `non-randomness' in rhythmic patterns by changing the null hypothesis. There are two ways of testing empirical data against a chosen null hypothesis: implementing a matching rhythm ratio transformation (Subsection~\ref{sub:alternative-ratio}), or weighting the data correspondingly (Subsection~\ref{sub:adjusting-normalization}).

For the first option (ratio rescaling), we can derive a new rhythm ratio formula, $s = f(q)$, which transforms the desired null distribution of intervals $p_I(i)$ into a uniform distribution of ratios $p_S(s) = 1$. For example, Figure~\ref{fig:ratio-distributions} demonstrates how a uniform distribution of intervals results in a non-uniform distribution of $r_k$ values. However, an appropriate choice of rescaled rhythm ratio formula $s = f(q)$ effectively "flattens" the rhythm ratio distribution; any subsequent statistical analysis can again correctly assume a uniform distribution for the rescaled ratio $s$, even though the null hypothesis is now different from a Poisson process.

For the second option (data normalization), we stick to the $r_k$ rhythm ratio formula and instead adapt the statistical analysis. The issue with a non-uniform null distribution is that not all observed ratio values are equally likely. For example, given the previously mentioned uniform distribution of intervals (Figure~\ref{fig:ratio-distributions}), there is a higher chance of getting an $r_k$ value close to 0.5 than to 0.2. So the expected number of $r_k$ values within a bin is not directly proportional to the width of the interval anymore; i.e., the normalization by bin width \parencite{roeske_categorical_2020} underestimates -- and thus undercorrects -- the expected number of points in a bin such as $[0.5, 0.6]$. Below, we show how to calculate the correct normalization factor. Note that this second option is most often the most straightforward one, and will be easier to practically apply in future research.

\subsection{Rescaling the rhythm ratio}
\label{sub:alternative-ratio}
Based on equations \ref{eq:p_q} and \ref{eq:p_s}, we can calculate the resulting probability distribution $p_{S}(s)$ of any rescaling $s = f(q)$ of the rhythm ratio $q$, for any null distribution of independently sampled intervals $p_I(i)$. Conversely, we can start from a given interval distribution $p_I(i)$ and a desired distribution of rhythm ratios. For example, a convenient choice is to impose that the variable $s$ (our new rhythm quantifier) is uniformly distributed over $[0, 1]$, i.e. $p_S(s) = 1$ for $0\le s \le 1,$ $p_S(s)=0$ otherwise. From these constraints, we can determine the corresponding ratio transformation function $f$ which achieves this goal.

To do so, we calculate the cumulative probability distribution of both distributions, $P_Q(q) = \int_0^q p_Q(x)\, dx$ and $P_{S}(s) = \int_0^q p_S(x)\, dx$, and combine them to get the two resulting transformation functions\footnote{The reason two resulting probability functions exist, $f_+$ and $f_-$, is that they correspond to one monotonically increasing and one monotonically decreasing function. $f_+$ is monotonically increasing and will map large values of $q = \frac{i_2}{i_1}$ (i.e., when $i_2 > i_1$) to large values of $s$, and a small $q$ to a small $s$ value. $f_-$ do the inverse: small values of $q$ will be mapped to large values of $s$, and as $q$ increases $f_-(q) = s$ will decrease.}, $f_+(q) = P_{S}^{-1}\left(P_Q(q)\right)$ and $f_-(q) = P_{S}^{-1}\left(1 - P_Q(q)\right)$. Since we have chosen $p_S(s) = 1$ for $s \in [0, 1],$ we simply have $P_S(s) = s$, in turn implying $P_S^{-1}(t) = t$. For independently sampled intervals with density $p_I(i)$, the cumulative distribution function of the ratio is
\begin{align}
    P_Q(q) &= \int_{0}^q p_Q(x)\, dx = \int_0^q \int_0^\infty t\, p_{I}(t)\, p_{I}(x\, t)\, dt\, dx \label{eq:P_q}
\end{align}

Consequently, the two possible ratio transformations resulting in a uniform probability density over $[0, 1]$ of ratios are
\begin{align}
    f_+(q) &= P_Q(q) = \int_0^q p_Q(x)\, dx \label{eq:f_+}
\end{align}
and
\begin{align}
    f_-(q) &= 1 - P_Q(q) = 1 - \int_0^q p_Q(x)\, dx \label{eq:f_-}
\end{align}
where $p_Q$ and $P_Q$ are calculated as in equations~\ref{eq:p_q} and \ref{eq:P_q}.

It can be helpful to note how these general formulas reduce to what we already know for a Poisson null hypothesis. For the specific case of a Poisson point process, with $p_I(i) = \lambda e^{-\lambda\, i}$ (equation~\ref{eq:p_i_poisson}), we know that $p_Q(q) = \frac{1}{(1 + q)^2}$ (equation~\ref{eq:p_q_poisson}). The two corresponding rhythm ratio formulas for a uniform ratio distribution over $[0, 1]$ (cfr. equations~\ref{eq:f_+} and \ref{eq:f_-}) are
\begin{align}
    f_+(q) &= \frac{q}{1 + q}
\end{align}
and
\begin{align}
    f_-(q) &= \frac{1}{1 + q} \label{eq:f_-_poisson}
\end{align}

Note that equation~\ref{eq:f_-_poisson} results in the familiar $r_k = f(q_k) = \frac{1}{1 + q_k} = \frac{i_k}{i_k + i_{k+ 1}}$ which indeed results in a uniform distribution of rhythm ratios\footnote{Also note that $r_k = \frac{i_k}{i_k + i_{k+1}}$ corresponds to the monotonically decreasing transformation function $f_-(q) = \frac{1}{1 + q}$: For example, when $q_k = \frac{i_{k+1}}{i_k}$ increases above 1, $r_k = \frac{i_k}{i_k + i_{k+1}}$ decreases below 0.5.}.

\subsection{Adjusting the normalization constant}
\label{sub:adjusting-normalization}
Finding a custom ratio transformation by solving equations~\ref{eq:f_+} and \ref{eq:f_-} can be non-trivial, particularly if one is interested in explicit formulas. As an alternative approach to the ratio rescaling of the previous section, it is possible to pick any rhythm ratio (such as $r_k$), and change the normalization or weight of the data points during the statistical analysis. For example, in the common binning approach (Section~\ref{sec:background}), we can normalize the bin counts differently. Note that the expected number of rhythm ratios that fall within a bin, under a specific null hypothesis, is proportional to the probability that a rhythm ratio $s$ falls into this bin. So instead of a division by the corresponding bin width, counts should be normalized by the total probability mass within a bin.

More formally, assume a null hypothesis distribution of interval durations $p_I$. To statistically test whether the number of rhythm ratios in two bins differs significantly \emph{with respect to this null hypothesis}, the count of rhythm ratios $s$ in a bin $b = [u, v]$ should be divided by a normalization factor $\hat{w}_{I, u, v}$:
\begin{align}
    \hat{w}_{I, u, v} &= \int_u^v p_{S}(s)\, ds \label{eq:w_I_uv}
\end{align}

with $p_{S}(s)$ being the resulting distribution of ratios $s$ (following equation~\ref{eq:p_s}). This normalization constant replaces the bin width in the normalization formula. For a bin $[u, v]$ which contains $m$ out of $N$ observed rhythm ratios, the normalized bin count then is $\frac{m}{N \cdot \hat{w}_{I, u, v}}$.

Note that equation~\ref{eq:w_I_uv} simplifies to $v - u$ (i.e., the width of bin $[u, v]$) for a uniform distribution $p_{S} = 1$. Therefore, under the Poisson process null hypothesis, this normalization factor equals the bin width, as commonly used in previous work \parencite[e.g.,][]{roeske_categorical_2020, de_gregorio_categorical_2021}. This shows how the commonly used normalization by bin width implicitly assumes a Poisson process as null hypothesis -- or more generally, it assumes that the distribution $p_S(s)$ is flat.

If the integral in equation \ref{eq:w_I_uv} is hard or impossible to solve analytically, it can be approximated numerically. In most cases, a simple Monte Carlo simulation suffices to approximate $\hat{w}_{I, u, v}$: Sample many pairs of (pseudo)random values from the null hypothesis distribution $p_{I}$, calculate the resulting ratios $s$ (e.g., $r_k$), count how many randomly sampled ratios fall within the $[u, v]$ bin, and calculate the corresponding fraction of the total number of samples; the resulting number is the normalization factor $\hat{w}_{I, u, v}$ (see Supporting Information, Algorithm~\ref{alg:normalization-pseudocode} for pseudocode and an example implementation in Python and R in Code fragments~\ref{lst:normalization-python} and \ref{lst:normalization-r}). For a large enough number of samples, the resulting value will approximate the normalization constant $\hat{w}_{I, u, v}$.

\begin{figure}[!h]
    \centering
    \includegraphics[width=\linewidth]{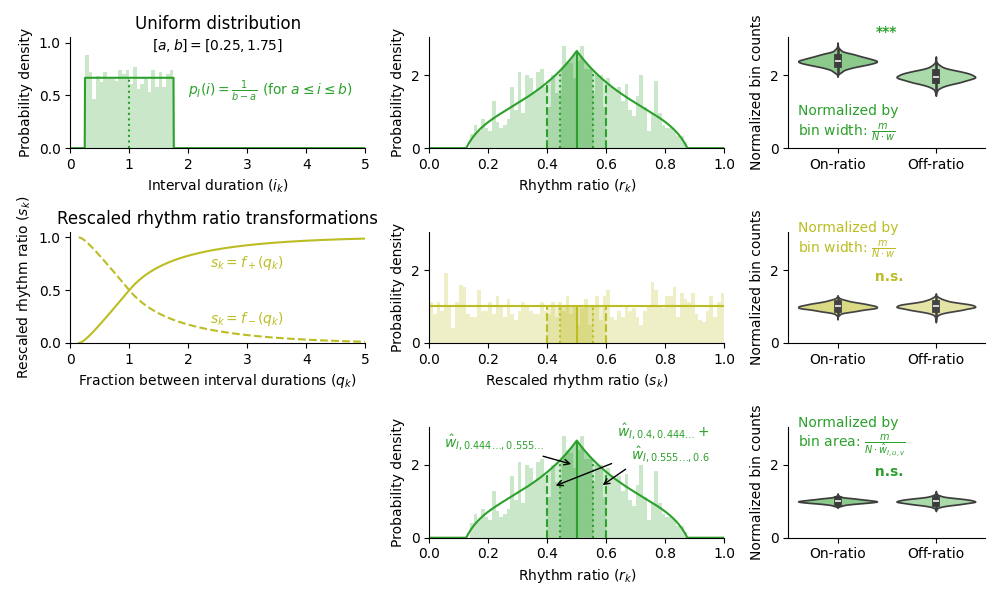}
    \caption{For a Poisson process null hypothesis, a uniform distribution of ratios results does not result in a uniform distribution of $r_k$ rhythm ratios (top row). In fact, a statistical comparison between 1000 sequences of 1000 intervals shows a significant different between on- and off-ratio bins for the 1:1 integer ratio (top right; Wilcoxon signed-rank test; $T = 515$, $p < 10^{-163}$; see Figure~\ref{fig:ratio-distributions}). A different rhythm ratio formula results in a uniform distribution of the rescaled rhythm ratio $s_k$ (middle row; for details, see Supplementary Information, Section~\ref{sec:full-example-uniform}). Given this rhythm ratio formula, there is no significant difference between the on- and off-ratio bins (Wilcoxon signed-rank test; $T = 244399$, $p \approx 0.52$). Alternatively, the bin count normalization of the original $r_k$ ratios can be changed (bottom row). If we do not divide by the bin width $w$, but instead by the area under the probability density curve within a bin ($\hat{w}_{I, u, v}$; equation~\ref{eq:w_I_uv}), statistical comparison is also non-significant (Wilcoxon signed-rank test; $T = 238765$, $p \approx 0.21$). Both non-significant results demonstrate how these approaches explicitly change the statistical null hypothesis.}
    \label{fig:uniform-ratio-distributions}
\end{figure}

If we apply this new normalization to the previous sample of uniformly distributed intervals, we can change the null hypothesis and test for statistically significant differences with respect to this uniform distribution. The high rate of statistically significant results we saw before is strongly reduced, confirming our derivation of the correct normalization factor. Both ways of adapting the null hypothesis, a rescaled rhythm ratio and a corrected normalization factor, are demonstrated in Figure~\ref{fig:uniform-ratio-distributions} for a uniform null distribution of intervals.

\section{An example: Testing alternative null hypotheses on birdsong and music data}
\label{sec:empirical-example}

Based on the above results, we now present a fully worked out example. We use four of the datasets originally analyzed by \textcite{roeske_categorical_2020} and test the empirical $r_k$ rhythm ratios against 2 new null hypotheses: We test which small-integer ratios are still significantly more often produced than chance if the integers would be 1) randomly sampled from a uniform distribution or 2) randomly sampled from a log-normal distribution. We start by deriving the respective null distributions of $r_k$ values, and then calculate the are under the curve in each of the on- and off-integer ratio bins. Once we have obtained these, we can correctly use them as normalization factors for the empirical data's bin counts.

\subsection{Uniformly distributed intervals}
\label{sub:uniform-intervals}
A practically relevant null hypothesis is a uniform distribution of interval durations. Several studies have simulated the resulting ratio distribution and compared their observed distribution using a Kolmogorov--Smirnov test \parencite{de_gregorio_categorical_2021, xing_syntactic_2022, anichini_measuring_2023, demartsev_male_2023}. Such an approach statistically compares the two distributions globally, but does not test the presence of peaks around specific integer ratios. As shown in Figure~\ref{fig:ratio-distributions}, if intervals are uniformly distributed, the resulting $r_k$ rhythm ratio distribution is more concentrated around 1:1 ($r_k = 0.5$). Consequently, the default normalization by bin width would \emph{underestimate} the null distribution's expected bin count in the on-ratio bins, and thus overestimate the concentration around the 1:1 ratio.

With the results and equations presented above, we can deduce the exact ratio distribution if intervals are uniformly distributed. Afterwards, we calculate the normalization factor to correctly test the bins around 1:1 to a uniform distribution. The full derivation of these formulas can be found in the Supporting Information (Section~\ref{sec:full-example-uniform}) and serve as an example of how to concretely apply our mathematical findings in other contexts. The unconcerned reader can skip the details of the following equations and continue reading the next subsection.

As intervals are distributed according to a uniform distribution over $[a, b]$, the probability density of the rhythm ratio $r$ (following equation~\ref{eq:p_r}; shown in the top middle pane in Figure~\ref{fig:uniform-ratio-distributions}) is
\begin{align}
p_R(r) = \begin{cases}
             \frac{1}{2 (b - a)^2} \left[\left(\frac{b}{1 - r}\right)^2 - \left(\frac{a}{r}\right)^2\right] & \text{if } \frac{a}{a + b} \leq r < \frac{1}{2} \\
             \frac{1}{2 (b - a)^2} \left[\left(\frac{b}{r}\right)^2 - \left(\frac{a}{1 - r}\right)^2\right] & \text{if } \frac{1}{2} \leq r \leq \frac{b}{a + b} \\
             0 & \text{otherwise}
         \end{cases} \label{eq:p_r_uniform}
\end{align}

To determine the total probability on an interval $[u, v]$ (i.e., the area under the probability density function between $u$ and $v$), we then need to calculate $P_R(v) - P_R(u)$, where $P_R(r)$ is the cumulative distribution function:
\begin{align}
P_R(r) = \begin{cases}
             0 & \text{if } 0 \leq r < \frac{a}{a + b} \\
             \frac{1}{2 (b - a)^2} \left[\frac{b^2}{1 - r} + \frac{a^2}{r} - (a + b)^2\right]  & \text{if } \frac{a}{a + b} \leq r < \frac{1}{2} \\
             1 - \frac{1}{2 (b - a)^2} \left[\frac{b^2}{r} + \frac{a^2}{1 - r} - (a + b)^2\right] & \text{if } \frac{1}{2} \leq r < \frac{b}{a + b} \\
             1 & \text{if } \frac{a}{a + b} \leq r \leq 1 \\
         \end{cases} \label{eq:P_r_uniform}
\end{align}

The new normalization factor for a bin $[u, v]$ can now be calculated as $P_R(v) - P_R(u)$. As a concrete example, assume a uniform distribution of interval durations between 0.25 and 1.75 (as in Figures~\ref{fig:ratio-distributions} and \ref{fig:uniform-ratio-distributions}): The resulting normalization factors for the on-ratio and off-ratio bins around the 1:1 ratio (as used by \textcite{roeske_categorical_2020} and others) are higher than the corresponding bin's widths:
\begin{align}
	\hat{w}_{U[0.25, 1.75], 0.444\ldots, 0.5} = \hat{w}_{U[0.25, 1.75], 0.5, 0.555\ldots} &\approx 0.13264 \\
	\hat{w}_{U[0.25, 1.75], 0.4, 0.444\ldots} = \hat{w}_{U[0.25, 1.75], 0.555\ldots, 0.6} &\approx 0.08727
\end{align}

Note how equations~\ref{eq:p_r_uniform} and \ref{eq:P_r_uniform} are specified in terms of $a$ and $b$. However, the absolute values of $a$ and $b$ do not influence the distribution of $r$; only the relative durations of the maximum and the minimum interval -- i.e., the fraction $\frac{b}{a}$ -- changes the shape of the ratio distribution\footnote{The interested reader can easily verify this by multiplying numerators and denominators of equations~\ref{eq:p_r_uniform} and \ref{eq:P_r_uniform} by $\frac{1}{a^2}$, naming $\frac{b}{a} = c$, and noticing that all occurrences of $a$ and $b$ can be replaced by $c$.}. For example, uniformly distributed intervals between \SI{10}{\ms} and \SI{50}{\ms} result in the same rhythm ratio distribution as intervals between \SI{3}{\s} and \SI{15}{\s}; in both cases the maximum interval is 5 times the minimum (i.e., $\frac{b}{a} = 5$), and the rhythm ratios ignore differences in average tempo. On the contrary, uniformly distributed intervals between \SI{10}{\ms} and \SI{40}{\ms} will result in a rhythm ratio distribution with a shape different from the above.

\subsection{Log-normally distributed intervals}
\label{sub:lognormal-intervals}
The second alternative null hypothesis we consider below is a log-normal distribution. Akin to a common normal distribution, a log-normal distribution naturally models a process producing interval durations are concentrated around a single value with added random noise. Crucially, the log-normal distribution has two theoretical advantages over the normal distribution: Measured interval durations cannot be negative, and randomly sampling from a log normal distribution will always produce a positive value. Moreover, many aspects of perception, including timing, are known to scale logarithmically \parencite{varshney_why_2013,grondin_about_2014}, and can thus be connected to a log-normal distribution.

We now derive the null distribution of $r_k$ values when interval durations are randomly sampled from a log-normal distribution. Once again, the unconcerned reader can skip the following equations and continue reading the following subsection.

A log-normal distribution is characterized by two parameters, $\mu$ and $\sigma$, respectively the mean and standard deviation in log-space. Under this parametrization, the null distribution of interval durations is:
\begin{align}
	p_I(i) = \frac{1}{i\, \sigma\sqrt{2\pi}}e^{-\frac{(\ln(i) - \mu)^2}{2\sigma^2}} \label{eq:p_i_lognormal}
\end{align}

The quotient of two log-normally distributed variables itself also follows a log-normal distribution, resulting in the following null probability density and cumulative distribution functions of the rhythm ratio $r$:
\begin{align}
	p_R(r) = \frac{1}{r\, (1 - r)\, 2\sigma\sqrt{\pi}}e^{-\frac{(\ln(1 - r) - \ln(r))^2}{4\sigma^2}} \label{eq:p_r_lognormal}
\end{align}

\begin{align}
	P_R(r) = \Phi\left(\frac{\ln(r) - \ln(1 - r)}{\sigma\sqrt{2}}\right) = \frac{1}{2} \left(1 + \erf\left(\frac{\ln(r) - \ln(1 - r)}{2\sigma}\right)\right) \label{eq:P_r_lognormal}
\end{align}

The corresponding normalization factor a different bin, $\hat{w}_{I, u, v}$, can be calculated by subtracting $P_R(v) - P_R(u)$. Taking for example $\sigma = 0.3$, as in Figure~\ref{fig:ratio-distributions}, the normalization factors are again larger than the bins' widths:
\begin{align}
	\hat{w}_{LN(\mu, 0.3), 0.444\ldots, 0.5} = \hat{w}_{LN(\mu, 0.3), 0.5, 0.555\ldots} &\approx 0.20054 \\
	\hat{w}_{LN(\mu, 0.3), 0.4, 0.444\ldots} = \hat{w}_{LN(\mu, 0.3), 0.555\ldots, 0.6} &\approx 0.12985
\end{align}

Finally, note that in equations~\ref{eq:p_r_lognormal} and \ref{eq:P_r_lognormal}, the log-normal distribution's scale parameter $\mu$ does not occur. Similar to the $\lambda$ parameter for the Poisson process and to the absolute scale of $a$ and $b$ for the uniform distribution, $\mu$ corresponds to the average interval duration and thus to the underlying tempo. As a result, $\mu$ the parameter cancels out in the calculations of $p_R(r)$.

\subsection{Rhythm ratios in birdsong and music}
We are now ready to reanalyze four of the datasets originally presented by \textcite{roeske_categorical_2020}, and explore how the alternative null hypotheses (i.e., uniformly or log-normally distributed interval durations) would influence the results. All data is openly available, and we refer to the previous study for exact details on recordings, onset annotations, and temporal intervals \parencite{roeske_categorical_2020, roeske_birdsong_2020, roeske_musical_2020}. We selected both songbird datasets, consisting of intervals extracted from recordings of thrush nightingale (\textit{Luscinia luscinia}) and zebra finch (\textit{Taeniopygia guttata}) songs. We also chose the Cuban salsa and Malian jembe as two distinct examples to test the presence of small-integer ratios in human music \parencite[part of the Interpersonal Entrainment in Music Performance (IEMP) corpus,][]{clayton_interpersonal_2018, clayton_interpersonal_2019}.

Starting from these datasets, we largely followed the same method of analysis as described by \textcite{roeske_categorical_2020}: We calculated the $r_k$ rhythm ratio between all pairs of adjacent intervals that had a longer "cycle length" (i.e., $i_k + i_{k + 1}$) than the thresholds determined in this previous study. To shortly explain these thresholds, \textcite{roeske_categorical_2020} found that interval pairs with a short combined duration mostly featured isochrony (dubbed "unimodal rhythms") and found most small-interval ratios to occur in pairs of intervals with a longer duration ("flexible rhythms"). We followed this approach and focused our analysis on the flexible rhythms, using the same thresholds. One minor difference to the previous study is that we also removed outliers by only using interval durations inside the 1st to 99th percentile of the data. This did not visibly affect the distribution of the $r_k$ values, but removed some very long intervals which were likely not part of the rhythmic structure and allowed us to better choose a fitting null hypothesis (see below).

We sorted the obtained $r_k$ values into the same bins as previous work; for example, the on-integer bin for the $2:1$ ratio (i.e., around $\frac{1}{3}$) contains all $r_k$ values between $\frac{1}{3.25}$ and $\frac{1}{2.75}$, while the two off-integer bins range from $\frac{1}{3.5}$ to $\frac{1}{3.25}$ and from $\frac{1}{2.75}$ to $\frac{1}{2.5}$. Following the approach by \textcite{roeske_categorical_2020}, we also combined the corresponding bin counts for the $1:2$ and $2:1$ ratios, and for the $1:3$ and $3:1$ ratios.

\begin{figure}[!h]
	\centering
	\includegraphics[width=\linewidth]{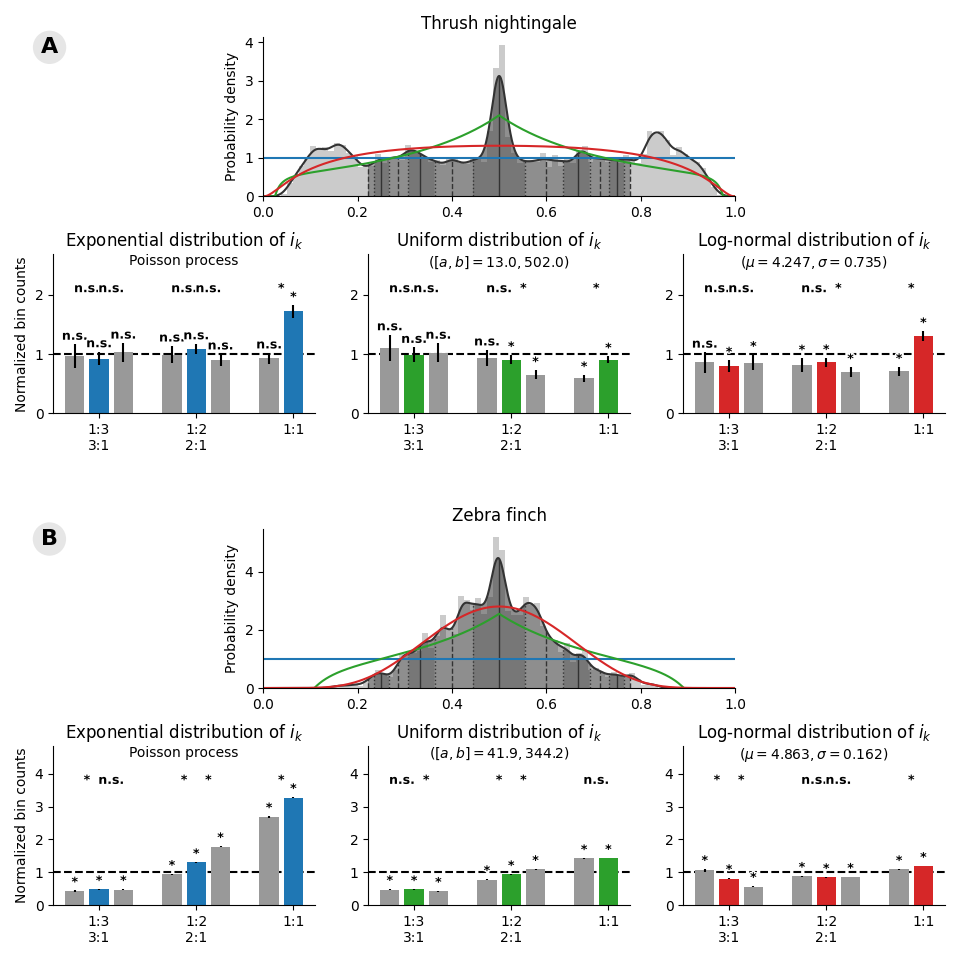}
	\caption{Our reanalysis of the thrush nightingale (\textbf{A}) and zebra finch (\textbf{B}) datasets \parencite{roeske_birdsong_2020} demonstrates the influence of the chose null hypothesis on the statistical results. We compare three selected null hypotheses: exponentially distributed (i.e., a Poisson process; blue), uniformly distributed (green), and log-normally distributed interval durations (red). The density plots present the estimated probability density of the observed $r_k$ ratios, and the expected distribution under each of the three null hypotheses. The bar chars present the bin frequencies, normalized according to each null hypothesis. The symbols on top (\textbf{*} and \textbf{n.s.}) show whether two adjacent on-/off-integer ratio bins contain a significantly different number of $r_k$ values. The statistical results right above each bar show whether a bin's normalized frequency is statistically significant from the null distribution.}
	\label{fig:empirical-examples-songbirds}
\end{figure}

\begin{figure}[!h]
	\centering
	\includegraphics[width=\linewidth]{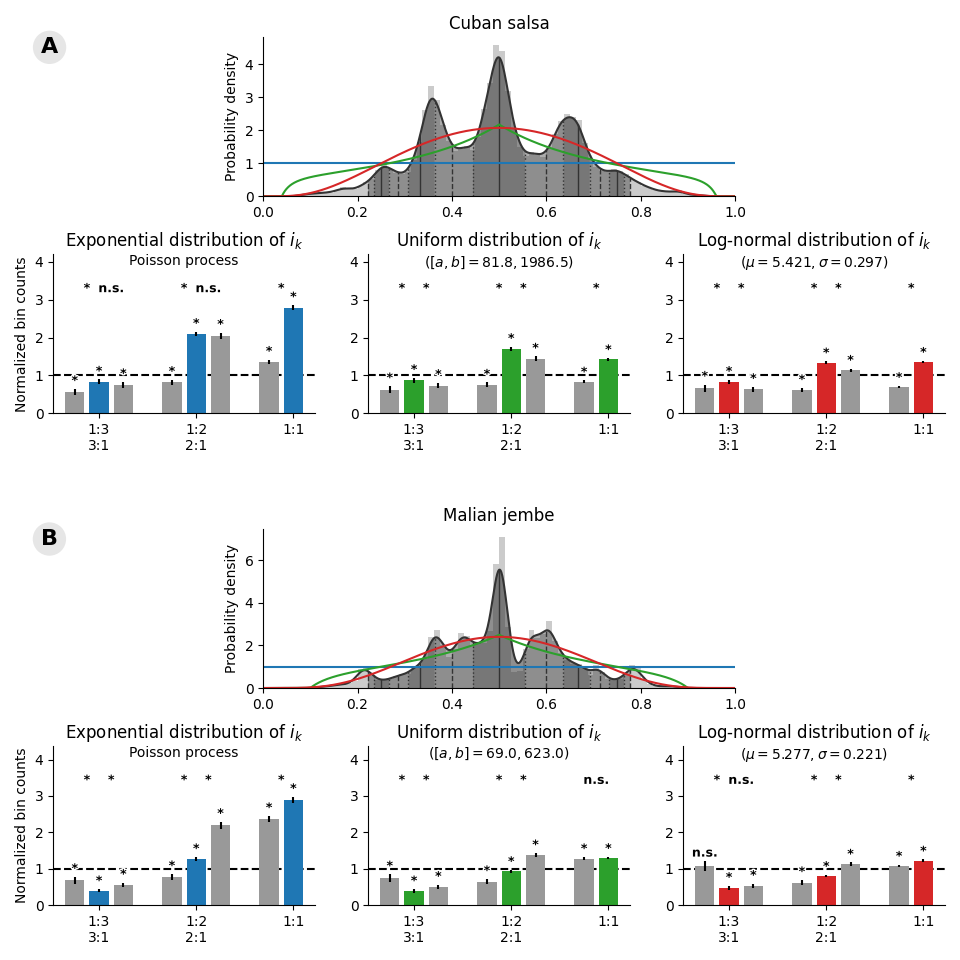}
	\caption{Our reanalysis of the Cuban salsa (\textbf{A}) and Malian jembe (\textbf{B}) datasets \parencite{roeske_musical_2020,clayton_interpersonal_2018} shows some potentially interesting differences between the different null hypotheses. The plots and their explanation can be read analogously to Figure~\ref{fig:empirical-examples-music}.}
	\label{fig:empirical-examples-music}
\end{figure}

A key step where our analyses differ from previous studies is the normalization of these bin counts. For our demonstration here, we chose three different null hypotheses against which to test the empirical $r_k$ distributions, which all resulted in a different normalization factor for each bin:
\begin{itemize}
	\item The first normalization is the one used in previous work \parencite[e.g.,][]{roeske_categorical_2020,de_gregorio_categorical_2021}. We divided the relative bin count (i.e., the number of $r_k$ values in a bin, divided by the overall total number of $r_k$ values) by the bins' width. This tests the empirical data with respect to the null hypothesis that intervals were generated by a Poisson process; i.e., the null distribution of the $i_k$ values is an exponential distribution.
	\item For the second normalization, we determined the expected distribution of $r_k$ values for interval durations uniformly sampled between the minimum and maximum observed duration (see Subsection~\ref{sub:uniform-intervals}).
	\item The third normalization assumed a log-normal distribution of interval durations. We fitted a log-normal distribution to the observed intervals, and calculated the expected $r_k$ distribution in this case (see Subsection~\ref{sub:lognormal-intervals}).
\end{itemize}
The $r_k$ distributions matching each of the three null hypotheses are shown in Figures~\ref{fig:empirical-examples-songbirds} and \ref{fig:empirical-examples-music}. For these cases, it was possible to derive the exact formulas to calculate the bins' normalization constant (see Sections~\ref{sub:uniform-intervals} and \ref{sub:lognormal-intervals}). Note however that we could have also approximated these constants numerically (see Subsection~\ref{sub:adjusting-normalization} for details). We provide an example of how to do so in Python and in R in the Supporting Information (Code fragments~\ref{lst:normalization-python} and \ref{lst:normalization-r}).

We performed the same bootstrapping procedure as described in detail by \textcite{roeske_categorical_2020}. In short, we resampled each $r_k$ distribution with replacement 1000 times in order to estimate the confidence intervals of each normalized bin frequency. Accordingly, when the Bonferroni-adjusted confidence intervals of an on- and off-integer ratio bin did not overlap, we considered the two normalized bin counts to be statistically significantly different. In addition to comparing on- and off-integer ratio bins, we also tested whether each bin in itself contained significantly more or less observations than expected by chance. To do so, we tested whether the confidence intervals of the normalized bin frequencies contained the value 1, the normalized bin frequency expected under the null hypothesis. These result indicate whether a bin contains significantly more or less $r_k$ values than the null hypothesis. The results of these statistical tests are shown in Figures~\ref{fig:empirical-examples-songbirds} and \ref{fig:empirical-examples-music}; a table with all results is included as Supporting Information (Tables~\ref{tab:results-thrush-nightingale}, \ref{tab:results-zebra-finch}, \ref{tab:results-cuban-salsa}, \ref{tab:results-malian-jembe}), as well as the Python code used to perform the analysis.

Figure~\ref{fig:empirical-examples-songbirds} shows the results of our approach for the two songbird corpora. Figure~\ref{fig:empirical-examples-music} shows the reanalysis of the Cuban salsa and Malian jembe datasets. One general observation is clear from these 4 examples: The uniform distribution (in blue) is typically a worse fit for the $r_k$ distributions than the two other null distributions. As a reminder, the two other null distributions (in green and red) are also achieved by sampling intervals randomly and independently; i.e., none of the null hypotheses have a specific mechanism which fine-tunes pairs of intervals to conform to small-integer ratios. As a result of this closer fit of the uniform and log-normal null hypotheses, most normalized bin counts end up much closer to 1 than for the Poisson process null hypothesis. These observations confirm our previous assertion that the Poisson process and its associated uniform $r_k$ distribution is a relatively low bar to statistically reject.

A full interpretation of our reanalysis would involve a critical assessment of which null hypotheses match the original research question, and would go beyond the scope of the current manuscript. We can however point out a couple of differences between the different analyses in Figures~\ref{fig:empirical-examples-songbirds} and \ref{fig:empirical-examples-music}. Note, for example, how the zebra finches' significant difference between the on- and off-$1:1$ bins disappears when tested against a null hypothesis of uniformly distributed intervals (Figure~\ref{fig:empirical-examples-songbirds}). This indicates that while the zebra finches produce more $r_k$ ratios close to $1:1$, this difference \emph{might} would not be unexpected if the birds would uniformly randomly sample intervals. On the contrary, the on-integer ratio bins around the $1:2$ and $1:3$ ratios in the Cuban salsa $r_k$ distribution do contain significantly more observed rhythm ratios than expected if interval durations were sampled uniformly or log-normally randomly (Figure~\ref{fig:empirical-examples-songbirds}).

Finally, we want to stress that these examples are meant to demonstrate our methodological findings regarding the often implicitly assumed null hypothesis in previous studies. They illustrate how to adapt an existing approach to a new, explicitly chosen null hypothesis. The rest of our approach follows a previous study as closely as possible \parencite[i.e., by][]{roeske_categorical_2020}. However, we have presented no arguments in favor or against following this exact approach. Other studies have approached the problem from a slightly different angle, have chosen different statistical tests, and their results are equally valid. Additionally, we intentionally do not discuss the choice of on- and off-integer bins: Our current findings have not provided any clear insights into the appropriate way to select boundaries for these bins. Altogether, the choice of null distribution is orthogonal to most of the other choices made when testing the presence of small-integer ratios in an empirical dataset. Our results do however highlight the importance of explicitly considering the choice of a null hypothesis, and demonstrate how this choice can be incorporated into any existing statistical approach.

\section{Conclusion and takeaway message}
\label{sec:conclusion}
The choice of a rhythm ratio formula has a strong effect on the analysis of temporal sequences: If we calculate the rhythm ratios of the same sequence with two different formulas, the resulting ratio distribution can be vastly different. The commonly used rhythm ratio $r_k$ formula, introduced by \textcite{roeske_categorical_2020}, results in a uniform distribution of ratios when applied to exponentially distributed intervals. Such an exponential distribution of intervals occurs between events generated by a homogeneous Poisson point process, a mathematical description of a highly random probabilistic temporal process which allows for events of arbitrarily small or large duration.

The mathematical relationship between a Poisson process, the rhythm ratio $r_k$ formula (equation \ref{eq:r_k}), and a uniform distribution of ratios means that the Poisson process often implicitly becomes the null hypothesis in statistical analyses. In several studies \parencite[e.g.,][]{roeske_categorical_2020, de_gregorio_categorical_2021, raimondi_isochrony_2023, lameira_recursive_2024}, an underlying assumption of the used statistical tests is a comparison to this uniform null distribution. A Poisson process may be a good default choice for a maximally random, arrhythmic baseline. We do however believe it is important to be aware of this implicit choice of null hypothesis.

For the case of biological organisms, a Poisson process is perhaps unrealistic: There is no limit on the duration of intervals, as intervals can get very close to 0 or very large. In other words, the null hypothesis from a Poisson process often sets the bar too low. We therefore advice that future research makes the choice of null hypothesis explicit. Doing so a) forces the researcher to consider whether this choice is appropriate, and b) allows that researcher to adapt the statistical analysis to the underlying scientific question.

Consequently, our findings also strengthen the case for the use of the $r_k$ rhythm ratio as originally presented by \textcite{roeske_categorical_2020}. The $r_k$ ratio is not an arbitrary or convenient choice, but represents a fundamental mathematical link to the random generation of temporal sequences. Given the intrinsic random properties of a Poisson process, the rhythm ratio formula is an appropriate default quantification of the relation between two intervals. Moreover, we also show how a sample of observed $r_k$ rhythm ratios can be weighted to statically test it against a more specific null hypothesis. As such, the $r_k$ ratio is as good as any other ratio to statistically analyze empirical data, but comes with the added benefit of being simple to calculate, symmetric around 1:1, and bounded between 0 and 1.

In conclusion, our takeaway message to future research into rhythmic categories and integer ratios is twofold. On the one hand, we recommend using the rhythm ratio $r_k$ to analyze the relationship between intervals and believe it is a good default choice. On the other hand, it is essential to remember to interpret any such results in the context of a Poisson process and carefully consider which other null hypothesis might be more appropriate within the context of the research question.

\section*{Acknowledgments}
The authors thank Jelle van der Werff, Dunia Giomo, and Vesta Eleuteri for their valuable feedback on the manuscript, Bart de Boer for additional mathematical input, and Christopher Fallaize for the book suggestions. The authors also thank three anonymous reviewers for their constructive feedback an suggestions.

YJ and AR are supported by the European Union (ERC, TOHR, 101041885) and by the HFSP research grant RGP0019/2022.

\section*{Author contributions}
\textbf{YJ}: Conceptualization, Methodology, Software, Formal analysis, Investigation, Writing - Original Draft, Visualization. \textbf{TT}: Methodology, Formal analysis, Investigation, Writing - Original Draft. \textbf{CC}: Software, Investigation. \textbf{MG}: Writing - Review \& Editing. \textbf{AR}: Conceptualization, Methodology, Investigation, Writing - Original Draft, Supervision, Funding acquisition.

\section*{Competing interest statement}
The authors have no competing interests to declare.

\printbibliography

\makeatletter\@input{supplement_aux.tex}\makeatother

\end{document}


\captionsetup[lstlisting]{singlelinecheck=false, margin=0pt, labelsep=space,labelfont=bf}
\renewcommand\lstlistingname{Code fragment}
\renewcommand\thesection{S\arabic{section}}
\renewcommand\thefigure{S\origthefigure}
\renewcommand\thelstlisting{S\arabic{lstlisting}}
\renewcommand\thealgorithm{S\arabic{algorithm}}
\renewcommand\thetable{S\arabic{table}}

\maketitle

\begin{algorithm}
\caption{Pseudocode to computationally approximate the normalization constant $w_{I, u, v}$. Parameters $u$ and $v$ are the left and right edge of the bin. Parameter $n$ is the number of samples used in the approximation; the higher $n$ the more accurate the estimate.} \label{alg:normalization-pseudocode}
\begin{algorithmic}

\Function{GetNormalizationConstant}{u, v, n}
    \State $m \gets 0$
    \For{$j = 1, \dots, n$}
        \State {$i_1 \gets $ \Call{SampleInterval}{}}
        \State {$i_2 \gets $ \Call{SampleInterval}{}}
        \State {$r \gets $ \Call{CalculateRhythmRatio}{$i_1$, $i_2$}} \Comment{Typically, $r \gets i_1 / (i_1 + i_2)$}
        \If{$u \le r \le v$}
            \State $m \gets m + 1$
        \EndIf
    \EndFor
    \State \Return $m / n$
\EndFunction

\end{algorithmic}
\end{algorithm}

\newpage
\begin{lstlisting}[language=Python, frame=lines, caption={Example Python implementation of the normalization algorithm in Algorithm~\ref{alg:normalization-pseudocode}. Concretely, this code calculates the normalization constant of the $r_k$ bin from 0.4 to 0.444\ldots for a uniform distribution of intervals. Adapt variables \texttt{u} and \texttt{v} to change the bin's edges, and increase \texttt{n} to get a more accurate approximation. Change \texttt{rng.uniform(...)} to sample from a different null distribution of intervals.}, label={lst:normalization-python}, keywordstyle=\ttfamily, showstringspaces=false]
import numpy as np

# Boundaries of the left off-ratio bin for 1:1
u, v = 1 / 2.5, 1 / 2.25

# Total number of samples to approximate 
n = 10**6

# Create a pseudorandom number generator object
rng = np.random.default_rng()

# Sample from a uniform distribution over [1, 5]
# Change to match the desired null distribution
intervals_1 = rng.uniform(1, 5, size=n)
intervals_2 = rng.uniform(1, 5, size=n)

ratios = intervals_1 / (intervals_1 + intervals_2)
bin_count = np.sum((u <= ratios) & (ratios <= v))
normalization_constant = bin_count / n

print("Normalization constant:", normalization_constant)
\end{lstlisting}

\newpage
\begin{lstlisting}[language=R, frame=lines, caption={Example R implementation of the normalization algorithm in Algorithm~\ref{alg:normalization-pseudocode}. Concretely, this code calculates the normalization constant of the $r_k$ bin from 0.4 to 0.444\ldots for a uniform distribution of intervals. Adapt variables \texttt{u} and \texttt{v} to change the bin's edges, and increase \texttt{n} to get a more accurate approximation. Change \texttt{runif(...)} to sample from a different null distribution of intervals.}, label={lst:normalization-r}, keywordstyle=\ttfamily, showstringspaces=false]
# Boundaries of the left off-ratio bin for 1:1
u <- 1 / 2.5
v <- 1 / 2.25

# Total number of samples to approximate 
n = 10^6

# Sample from a uniform distribution over [1, 5]
# Change to match the desired null distribution
intervals_1 <- runif(n, 1, 5)
intervals_2 <- runif(n, 1, 5)

ratios <- intervals_1 / (intervals_1 + intervals_2)
bin_count <- sum((u <= ratios) & (ratios <= v))
normalization_constant <- bin_count / n

paste("Normalization constant:", normalization_constant)
\end{lstlisting}

\newpage

\begin{table}[h]
	\centering
	\resizebox{\textwidth}{!}{\begin{tabular}{lrrlrrrr}
\toprule
\makecell{\textbf{Null-hypothesis}} & \makecell{\textbf{Bin (lower)}} & \makecell{\textbf{Bin (upper)}} & \makecell{\textbf{On-/off-}\\\textbf{integer}} & \makecell{\textbf{Bin count}} & \makecell{\textbf{Normalized}\\\textbf{frequency}} & \makecell{\textbf{Normalized}\\\textbf{CI (lower)}} & \makecell{\textbf{Normalized}\\\textbf{CI (upper)}} \\
\midrule
Poisson process & 0.222222 & 0.235294 & Off & 159 & 0.971836 & 0.770104 & 1.167487 \\
Poisson process & 0.235294 & 0.266667 & On & 363 & 0.924467 & 0.814944 & 1.041642 \\
Poisson process & 0.266667 & 0.285714 & Off & 245 & 1.027685 & 0.872462 & 1.187102 \\
Poisson process & 0.285714 & 0.307692 & Off & 272 & 0.988814 & 0.846981 & 1.137864 \\
Poisson process & 0.307692 & 0.363636 & On & 758 & 1.082554 & 0.994001 & 1.175414 \\
Poisson process & 0.363636 & 0.400000 & Off & 409 & 0.898650 & 0.790944 & 1.004170 \\
Poisson process & 0.400000 & 0.444444 & Off & 517 & 0.929410 & 0.830528 & 1.021111 \\
Poisson process & 0.444444 & 0.500000 & On & 1195 & 1.718600 & 1.612169 & 1.823600 \\
Uniform & 0.222222 & 0.235294 & Off & 159 & 1.105373 & 0.875921 & 1.327908 \\
Uniform & 0.235294 & 0.266667 & On & 363 & 0.989779 & 0.872519 & 1.115233 \\
Uniform & 0.266667 & 0.285714 & Off & 245 & 1.026315 & 0.871299 & 1.185519 \\
Uniform & 0.285714 & 0.307692 & Off & 272 & 0.931467 & 0.797860 & 1.071872 \\
Uniform & 0.307692 & 0.363636 & On & 758 & 0.907492 & 0.833258 & 0.985335 \\
Uniform & 0.363636 & 0.400000 & Off & 409 & 0.652308 & 0.574127 & 0.728902 \\
Uniform & 0.400000 & 0.444444 & Off & 517 & 0.588671 & 0.526041 & 0.646753 \\
Uniform & 0.444444 & 0.500000 & On & 1195 & 0.906728 & 0.850575 & 0.962125 \\
Log-normal & 0.222222 & 0.235294 & Off & 159 & 0.861263 & 0.682483 & 1.034653 \\
Log-normal & 0.235294 & 0.266667 & On & 363 & 0.793556 & 0.699543 & 0.894139 \\
Log-normal & 0.266667 & 0.285714 & Off & 245 & 0.856353 & 0.727009 & 0.989193 \\
Log-normal & 0.285714 & 0.307692 & Off & 272 & 0.808110 & 0.692197 & 0.929921 \\
Log-normal & 0.307692 & 0.363636 & On & 758 & 0.860421 & 0.790038 & 0.934226 \\
Log-normal & 0.363636 & 0.400000 & Off & 409 & 0.697910 & 0.614264 & 0.779859 \\
Log-normal & 0.400000 & 0.444444 & Off & 517 & 0.712881 & 0.637036 & 0.783218 \\
Log-normal & 0.444444 & 0.500000 & On & 1195 & 1.308058 & 1.227051 & 1.387975 \\
\bottomrule
\end{tabular}
}
	\caption{Full results of the thrush nightingale dataset (main text, Figure~\ref{fig:empirical-examples-songbirds}A), presenting the bin count, normalized frequency, and normalized confidence interval (CI) for each combination of bin and null-hypothesis.}
	\label{tab:results-thrush-nightingale}
\end{table}

\newpage

\begin{table}[h]
	\centering
	\resizebox{\textwidth}{!}{\begin{tabular}{lrrlrrrr}
\toprule
\makecell{\textbf{Null-hypothesis}} & \makecell{\textbf{Bin (lower)}} & \makecell{\textbf{Bin (upper)}} & \makecell{\textbf{On-/off-}\\\textbf{integer}} & \makecell{\textbf{Bin count}} & \makecell{\textbf{Normalized}\\\textbf{frequency}} & \makecell{\textbf{Normalized}\\\textbf{CI (lower)}} & \makecell{\textbf{Normalized}\\\textbf{CI (upper)}} \\
\midrule
Poisson process & 0.222222 & 0.235294 & Off & 2998 & 0.428371 & 0.407936 & 0.448092 \\
Poisson process & 0.235294 & 0.266667 & On & 8015 & 0.477178 & 0.463305 & 0.490638 \\
Poisson process & 0.266667 & 0.285714 & Off & 4814 & 0.472054 & 0.454403 & 0.490592 \\
Poisson process & 0.285714 & 0.307692 & Off & 10975 & 0.932701 & 0.910945 & 0.956157 \\
Poisson process & 0.307692 & 0.363636 & On & 38938 & 1.300008 & 1.283714 & 1.313799 \\
Poisson process & 0.363636 & 0.400000 & Off & 34621 & 1.778275 & 1.755516 & 1.801396 \\
Poisson process & 0.400000 & 0.444444 & Off & 63944 & 2.687255 & 2.664308 & 2.711337 \\
Poisson process & 0.444444 & 0.500000 & On & 97432 & 3.275674 & 3.254020 & 3.296454 \\
Uniform & 0.222222 & 0.235294 & Off & 2998 & 0.472742 & 0.450190 & 0.494505 \\
Uniform & 0.235294 & 0.266667 & On & 8015 & 0.475841 & 0.462006 & 0.489263 \\
Uniform & 0.266667 & 0.285714 & Off & 4814 & 0.424731 & 0.408850 & 0.441411 \\
Uniform & 0.285714 & 0.307692 & Off & 10975 & 0.776241 & 0.758134 & 0.795762 \\
Uniform & 0.307692 & 0.363636 & On & 38938 & 0.938307 & 0.926546 & 0.948261 \\
Uniform & 0.363636 & 0.400000 & Off & 34621 & 1.089860 & 1.075912 & 1.104031 \\
Uniform & 0.400000 & 0.444444 & Off & 63944 & 1.421378 & 1.409241 & 1.434116 \\
Uniform & 0.444444 & 0.500000 & On & 97432 & 1.430219 & 1.420765 & 1.439292 \\
Log-normal & 0.222222 & 0.235294 & Off & 2998 & 1.054677 & 1.004366 & 1.103232 \\
Log-normal & 0.235294 & 0.266667 & On & 8015 & 0.807426 & 0.783951 & 0.830201 \\
Log-normal & 0.266667 & 0.285714 & Off & 4814 & 0.564152 & 0.543058 & 0.586306 \\
Log-normal & 0.285714 & 0.307692 & Off & 10975 & 0.876676 & 0.856226 & 0.898723 \\
Log-normal & 0.307692 & 0.363636 & On & 38938 & 0.849461 & 0.838813 & 0.858472 \\
Log-normal & 0.363636 & 0.400000 & Off & 34621 & 0.857888 & 0.846909 & 0.869042 \\
Log-normal & 0.400000 & 0.444444 & Off & 63944 & 1.091312 & 1.081993 & 1.101092 \\
Log-normal & 0.444444 & 0.500000 & On & 97432 & 1.192733 & 1.184849 & 1.200300 \\
\bottomrule
\end{tabular}
}
	\caption{Full results of the zebra finch dataset (main text, Figure~\ref{fig:empirical-examples-songbirds}B), presenting the bin count, normalized frequency, and normalized confidence interval (CI) for each combination of bin and null-hypothesis.}
	\label{tab:results-zebra-finch}
\end{table}

\newpage

\begin{table}[h]
	\centering
	\resizebox{\textwidth}{!}{\begin{tabular}{lrrlrrrr}
\toprule
\makecell{\textbf{Null-hypothesis}} & \makecell{\textbf{Bin (lower)}} & \makecell{\textbf{Bin (upper)}} & \makecell{\textbf{On-/off-}\\\textbf{integer}} & \makecell{\textbf{Bin count}} & \makecell{\textbf{Normalized}\\\textbf{frequency}} & \makecell{\textbf{Normalized}\\\textbf{CI (lower)}} & \makecell{\textbf{Normalized}\\\textbf{CI (upper)}} \\
\midrule
Poisson process & 0.222222 & 0.235294 & Off & 340 & 0.564037 & 0.477772 & 0.643666 \\
Poisson process & 0.235294 & 0.266667 & On & 1205 & 0.832922 & 0.775547 & 0.893767 \\
Poisson process & 0.266667 & 0.285714 & Off & 658 & 0.749122 & 0.678536 & 0.827683 \\
Poisson process & 0.285714 & 0.307692 & Off & 829 & 0.817962 & 0.749876 & 0.887035 \\
Poisson process & 0.307692 & 0.363636 & On & 5394 & 2.090856 & 2.027654 & 2.152491 \\
Poisson process & 0.363636 & 0.400000 & Off & 3437 & 2.049649 & 1.969112 & 2.128385 \\
Poisson process & 0.400000 & 0.444444 & Off & 2779 & 1.355933 & 1.292503 & 1.417419 \\
Poisson process & 0.444444 & 0.500000 & On & 7127 & 2.781932 & 2.714792 & 2.851806 \\
Uniform & 0.222222 & 0.235294 & Off & 340 & 0.628959 & 0.532765 & 0.717753 \\
Uniform & 0.235294 & 0.266667 & On & 1205 & 0.872044 & 0.811975 & 0.935747 \\
Uniform & 0.266667 & 0.285714 & Off & 658 & 0.730003 & 0.661219 & 0.806559 \\
Uniform & 0.285714 & 0.307692 & Off & 829 & 0.750884 & 0.688381 & 0.814292 \\
Uniform & 0.307692 & 0.363636 & On & 5394 & 1.705079 & 1.653538 & 1.755341 \\
Uniform & 0.363636 & 0.400000 & Off & 3437 & 1.445373 & 1.388580 & 1.500896 \\
Uniform & 0.400000 & 0.444444 & Off & 2779 & 0.833694 & 0.794694 & 0.871498 \\
Uniform & 0.444444 & 0.500000 & On & 7127 & 1.423875 & 1.389511 & 1.459639 \\
Log-normal & 0.222222 & 0.235294 & Off & 340 & 0.666431 & 0.564506 & 0.760515 \\
Log-normal & 0.235294 & 0.266667 & On & 1205 & 0.827581 & 0.770574 & 0.888035 \\
Log-normal & 0.266667 & 0.285714 & Off & 658 & 0.632040 & 0.572486 & 0.698322 \\
Log-normal & 0.285714 & 0.307692 & Off & 829 & 0.617420 & 0.566027 & 0.669559 \\
Log-normal & 0.307692 & 0.363636 & On & 5394 & 1.336372 & 1.295976 & 1.375766 \\
Log-normal & 0.363636 & 0.400000 & Off & 3437 & 1.137538 & 1.092841 & 1.181236 \\
Log-normal & 0.400000 & 0.444444 & Off & 2779 & 0.695322 & 0.662796 & 0.726852 \\
Log-normal & 0.444444 & 0.500000 & On & 7127 & 1.356875 & 1.324127 & 1.390955 \\
\bottomrule
\end{tabular}
}
	\caption{Full results of the Cuban salsa finch dataset (main text, Figure~\ref{fig:empirical-examples-music}A), presenting the bin count, normalized frequency, and normalized confidence interval (CI) for each combination of bin and null-hypothesis.}
	\label{tab:results-cuban-salsa}
\end{table}

\newpage

\begin{table}[h]
	\centering
	\resizebox{\textwidth}{!}{\begin{tabular}{lrrlrrrr}
\toprule
\makecell{\textbf{Null-hypothesis}} & \makecell{\textbf{Bin (lower)}} & \makecell{\textbf{Bin (upper)}} & \makecell{\textbf{On-/off-}\\\textbf{integer}} & \makecell{\textbf{Bin count}} & \makecell{\textbf{Normalized}\\\textbf{frequency}} & \makecell{\textbf{Normalized}\\\textbf{CI (lower)}} & \makecell{\textbf{Normalized}\\\textbf{CI (upper)}} \\
\midrule
Poisson process & 0.222222 & 0.235294 & Off & 332 & 0.678619 & 0.590705 & 0.774699 \\
Poisson process & 0.235294 & 0.266667 & On & 459 & 0.390921 & 0.346630 & 0.438620 \\
Poisson process & 0.266667 & 0.285714 & Off & 393 & 0.551288 & 0.486761 & 0.618628 \\
Poisson process & 0.285714 & 0.307692 & Off & 630 & 0.765911 & 0.685667 & 0.852234 \\
Poisson process & 0.307692 & 0.363636 & On & 2649 & 1.265187 & 1.206914 & 1.321074 \\
Poisson process & 0.363636 & 0.400000 & Off & 2983 & 2.191859 & 2.102924 & 2.288123 \\
Poisson process & 0.400000 & 0.444444 & Off & 3942 & 2.369877 & 2.280300 & 2.447433 \\
Poisson process & 0.444444 & 0.500000 & On & 6016 & 2.893390 & 2.818843 & 2.972756 \\
Uniform & 0.222222 & 0.235294 & Off & 332 & 0.741848 & 0.645743 & 0.846879 \\
Uniform & 0.235294 & 0.266667 & On & 459 & 0.389394 & 0.345275 & 0.436906 \\
Uniform & 0.266667 & 0.285714 & Off & 393 & 0.498755 & 0.440377 & 0.559678 \\
Uniform & 0.285714 & 0.307692 & Off & 630 & 0.643388 & 0.575980 & 0.715902 \\
Uniform & 0.307692 & 0.363636 & On & 2649 & 0.926259 & 0.883596 & 0.967174 \\
Uniform & 0.363636 & 0.400000 & Off & 2983 & 1.367593 & 1.312103 & 1.427656 \\
Uniform & 0.400000 & 0.444444 & Off & 3942 & 1.278743 & 1.230409 & 1.320591 \\
Uniform & 0.444444 & 0.500000 & On & 6016 & 1.290888 & 1.257629 & 1.326297 \\
Log-normal & 0.222222 & 0.235294 & Off & 332 & 1.062150 & 0.924550 & 1.212530 \\
Log-normal & 0.235294 & 0.266667 & On & 459 & 0.473538 & 0.419886 & 0.531317 \\
Log-normal & 0.266667 & 0.285714 & Off & 393 & 0.525177 & 0.463706 & 0.589328 \\
Log-normal & 0.285714 & 0.307692 & Off & 630 & 0.618870 & 0.554031 & 0.688621 \\
Log-normal & 0.307692 & 0.363636 & On & 2649 & 0.798110 & 0.761350 & 0.833365 \\
Log-normal & 0.363636 & 0.400000 & Off & 2983 & 1.122131 & 1.076600 & 1.171413 \\
Log-normal & 0.400000 & 0.444444 & Off & 3942 & 1.078591 & 1.037822 & 1.113889 \\
Log-normal & 0.444444 & 0.500000 & On & 6016 & 1.222216 & 1.190726 & 1.255742 \\
\bottomrule
\end{tabular}
}
	\caption{Full results of the Malian jembe finch dataset (main text, Figure~\ref{fig:empirical-examples-music}B), presenting the bin count, normalized frequency, and normalized confidence interval (CI) for each combination of bin and null-hypothesis.}
	\label{tab:results-malian-jembe}
\end{table}

\newpage

\section{Full derivation of probability distributions for uniformly distributed intervals}
\label{sec:full-example-uniform}

Assume intervals are distributed uniformly over $[a, b]$. I.e., the probability density of intervals $p_I$ is:

\begin{align}
    p_I(i) = 
    \begin{cases}
        \frac{1}{b - a} & \text{if } a \leq i \leq b \\
        0 & \text{otherwise}
    \end{cases}
\end{align}

To determine the distribution $p_Q$ of $q = \frac{i_2}{i_1}$, we calculate the following integral:

\begin{align}
    p_Q(q) = \int_0^\infty t\, p_I(t)\, p_I(q\, t)\, dt
\end{align}

If $0 < q < \frac{a}{b}$ or $q > \frac{b}{a}$, there is no way to have both $a \leq t \leq b$ and $a \leq q\, t \leq b$.

\begin{itemize}
    \item If $0 < q < \frac{a}{b}$:
          \begin{align*}
              a \leq q\, t < \frac{a}{b} t &\Rightarrow t > \frac{b}{a} a = b \\
              t \leq b &\Rightarrow q\, t \leq q\, b < \frac{a}{b} b = a
          \end{align*}
    \item If $q > \frac{b}{a}$:
          \begin{align*}
              b \geq q\, t > \frac{b}{a} t &\Rightarrow t < \frac{a}{b} b = a \\
              t \geq a &\Rightarrow q\, t \geq q\, a > \frac{b}{a} a = b
          \end{align*}
\end{itemize}

So, for $0 < q < \frac{a}{b}$ or $q > \frac{b}{a}$, either $p_I(t) = 0$ or $p_I(q\, t) = 0$ and consequently $p_Q(q) = 0$. So we just need to calculate $p_Q(q)$ for $\frac{a}{b} \leq q \leq \frac{b}{a}$. There are again two cases:

\begin{itemize}
    \item If $\frac{a}{b} \leq q \leq 1$:
          \begin{align*}
              p_I(t) > 0 &\Leftrightarrow a \leq t \leq b \\
              p_I(q\, t) > 0 &\Leftrightarrow \frac{a}{q} \leq t \leq \frac{b}{q} \\
              p_I(t)\, p_I(q\, t) > 0 &\Leftrightarrow \frac{a}{q} \leq t \leq b
          \end{align*}

          So, we can reduce the interval of integration to $[\frac{a}{q}, b]$:
          \begin{align*}
              p_Q(q) &= \int_{\frac{a}{q}}^b t \left(\frac{1}{b - a}\right) \left(\frac{1}{b - a}\right)\, dt \\
                     &= \frac{1}{(b - a)^2} \int_{\frac{a}{q}}^b t\, dt \\
                     &= \frac{1}{2 (b - a)^2} \left[b^2 - \left(\frac{a}{q}\right)^2\right]
          \end{align*}

    \item Symmetrically, if $1 \leq q \leq \frac{b}{a}$:
          \begin{align*}
               p_I(t)\, p_I(q\, t) > 0 &\Leftrightarrow a \leq t \leq \frac{b}{q}
          \end{align*}
          and
          \begin{align*}
              p_Q(q) &= \int_a^{\frac{b}{q}} t \left(\frac{1}{b - a}\right) \left(\frac{1}{b - a}\right)\, dt \\
                     &= \frac{1}{(b - a)^2} \int_a^{\frac{b}{q}} t\, dt \\
                     &= \frac{1}{2 (b - a)^2} \left[\left(\frac{b}{q}\right)^2 - a^2\right]
          \end{align*}
\end{itemize}

Altogether, $p_Q(q)$ becomes
\begin{align}
    p_Q(q) = \begin{cases}
                    \frac{1}{2 (b - a)^2} \left[b^2 - \left(\frac{a}{q}\right)^2\right] & \text{if } \frac{a}{b} \leq q < 1 \\
                    \frac{1}{2 (b - a)^2} \left[\left(\frac{b}{q}\right)^2 - a^2\right] & \text{if } 1 \leq q \leq \frac{b}{a} \\
                    0 & \text{otherwise}
                \end{cases}
\end{align}

The cumulative probability distribution of $q$ is
\begin{align}
    P_Q(q) &= \int_0^q p_Q(t)\, dt \nonumber \\
           &= \begin{cases}
                  0 & \text{if } 0 \leq q < \frac{a}{b} \\
                  \int_\frac{a}{b}^q p_Q(t)\, dt & \text{if } \frac{a}{b} \leq q < 1 \\
                  1 - \int_q^\frac{b}{a} p_Q(t)\, dt & \text{if } 1 \leq q < \frac{b}{a} \\
                  1 & \text{if } \frac{b}{a} \leq q
              \end{cases} \nonumber \\
           &= \begin{cases}
                  0 & \text{if } 0 \leq q < \frac{a}{b} \\
                  \frac{1}{2 (b - a)^2} \left[qb^2 + \frac{a^2}{q} - 2ab\right] & \text{if } \frac{a}{b} \leq q < 1 \\
                  1 - \frac{1}{2 (b - a)^2} \left[\frac{b^2}{q} + qa^2 - 2ab\right] & \text{if } 1 \leq q < \frac{b}{a} \\
                  1 & \text{if } \frac{b}{a} \leq q
              \end{cases}
\end{align}

This formulation of $P_Q(q)$ then determines the two rescaling formulas $f_+(q)$ and $f_-(q)$, that map a uniform distribution of intervals into uniform distribution of ratios $s$:
\begin{align}
    f_+(q) &= P_Q(q) \\
    f_-(q) &= 1 - P_Q(q)
\end{align}

Performing the change of variable according to ratio transformation $r = f(q) = \frac{1}{1 + q}$ or $q = f^{-1}(r) = \frac{1}{r} - 1$, we get probability distribution $p_R(r)$, as plotted in the main manuscript's figures:
\begin{align}
    p_R(r) &= p_Q\left(\frac{1}{r} - 1\right) \left| \frac{d}{dr}\left(\frac{1}{r} - 1\right) \right| \nonumber \\
           &= \frac{1}{r^2}\, p_Q²\left(\frac{1 - r}{r}\right) \nonumber \\
           &= \begin{cases}
                  \frac{1}{2 (b - a)^2} \left[\left(\frac{b}{1 - r}\right)^2 - \left(\frac{a}{r}\right)^2\right] & \text{if } \frac{a}{a + b} \leq r < \frac{1}{2} \\
                  \frac{1}{2 (b - a)^2} \left[\left(\frac{b}{r}\right)^2 - \left(\frac{a}{1 - r}\right)^2\right] & \text{if } \frac{1}{2} \leq r \leq \frac{b}{a + b} \\
                  0 & \text{otherwise}
              \end{cases}
\end{align}

To get the total probability in a bin $[u, v]$, we integrate $p_R(r)$, assuming $\frac{a}{a + b} \leq u \leq v \leq \frac{b}{a + b}$:
\begin{align}
    \int_u^v p_R(r) &= \begin{cases}
                           \frac{1}{2 (b - a)^2} \left[\frac{b^2}{1 - v} + \frac{a^2}{v} - \frac{b^2}{1 - u} - \frac{a^2}{u}\right] & \text{if } \frac{a}{a + b} \leq u \leq v < \frac{1}{2} \\
                           \frac{1}{2 (b - a)^2} \left[-\frac{b^2}{v} - \frac{a^2}{1 - v} + \frac{b^2}{u} + \frac{a^2}{1 - u}\right] & \text{if } \frac{1}{2} \leq u \leq v \leq  \frac{b}{a + b} \\
                           \frac{1}{2 (b - a)^2} \left[4b^2 + 4a^2 - \frac{b^2}{1 - u} - \frac{a^2}{u}-\frac{b^2}{v} - \frac{a^2}{1 - v}\right] & \text{if } \frac{a}{a + b} \leq u \leq \frac{1}{2} \leq v \leq \frac{b}{a + b}
                       \end{cases}
\end{align}

An alternative, more elegant way to describe $\int_u^v p_R(r)\, dr$ is to calculate the cumulative probability distribution $P_R(r)$:
\begin{align}
    P_R(r) &= \int_0^r p_R(t)\, dt \nonumber \\
           &= \begin{cases}
                  0 & \text{if } 0 \leq r < \frac{a}{a + b} \\
                  \int_\frac{a}{a + b}^r p_R(t)\, dt & \text{if } \frac{a}{a + b} \leq r < \frac{1}{2} \\
                  1 - \int_\frac{1}{2}^\frac{b}{a + b} p_R(t)\, dt & \text{if } \frac{1}{2} \leq r < \frac{b}{a + b} \\
                  1 & \text{if } \frac{a}{a + b} \leq r \leq 1 \\
              \end{cases} \nonumber \\
           &= \begin{cases}
                  0 & \text{if } 0 \leq r < \frac{a}{a + b} \\
                  \frac{1}{2 (b - a)^2} \left[\frac{b^2}{1 - r} + \frac{a^2}{r} - (a + b)^2\right]  & \text{if } \frac{a}{a + b} \leq r < \frac{1}{2} \\
                  1 - \frac{1}{2 (b - a)^2} \left[\frac{b^2}{r} + \frac{a^2}{1 - r} - (a + b)^2\right] & \text{if } \frac{1}{2} \leq r < \frac{b}{a + b} \\
                  1 & \text{if } \frac{a}{a + b} \leq r \leq 1 \\
              \end{cases}
\end{align}

Now, the area under the curve in interval $[u, v]$ is equal to $P_R(v) - P_R(u)$.

For example, in the case of the bins used by \textcite{roeske_categorical_2020} and others, these are the normalization constants $\hat{w}_{I, u, v}$ for, respectively, the on-ratio and off-ratio bins around the 1:1 ratio:
\begin{align}
	\hat{w}_{I, 0.444\ldots, 0.5} = \hat{w}_{I, 0.5, 0.555\ldots} &= \frac{1}{2} - \frac{(4b - 5a)^2}{40 (b - a)^2} \\
	\hat{w}_{I, 0.4, 0.444\ldots} = \hat{w}_{I, 0.555\ldots, 0.6} &= \frac{(4b - 5a)^2}{40 (b - a)^2} - \frac{(2b - 3a)^2}{12 (b - a)^2}
\end{align}

\printbibliography

\makeatletter\@input{main_aux.tex}\makeatother